\documentclass[12pt,letterpaper]{article}

\pdfoutput=1

\usepackage[margin=1in]{geometry}
\usepackage[T1]{fontenc}
\usepackage[utf8]{inputenc}
\usepackage{amsmath,amssymb}
\usepackage{mathtools}
\usepackage{physics}
\usepackage{dsfont}
\usepackage{tikz}
\usepackage{float}
\numberwithin{equation}{section}

\usepackage{authblk}

\usepackage{graphicx}
\usepackage{xcolor}
\usepackage[numbers,sort&compress]{natbib}
\usepackage[colorlinks=true,linkcolor=blue,citecolor=blue,urlcolor=magenta]{hyperref}

\newcommand{\defeq}{\vcentcolon=}
\newcommand{\ODF}{\Omega^{\mathrm{DF}}_{\mathcal{R}}}
\newcommand{\spl}[1]{\mathrm{SL}(#1,\mathbb{R})}
\newcommand{\updown}[2]{^{#1}_{\phantom{#1}#2}}
\newcommand{\downup}[2]{_{#1}^{\phantom{#1}#2}}

\newcommand{\ldv}[1]{\mathcal{L}_{#1}}

\newcommand{\hg}{\hat{g}}

\DeclareMathOperator*{\doublewedge}{\,\wedge\mkern-15mu\wedge\,}
\makeatletter
\newenvironment{Align}
  {\begin{equation}\begin{aligned}}
  {\end{aligned}\end{equation}\@endpetrue\ignorespacesafterend}
\makeatother
\usepackage{tocloft}

\title{Black Hole Entropy from Entangled Quantum Reference Frames}
\author[1]{Ludovic Varrin\thanks{ludovic.varrin@cpt.univ-mrs.fr}}
\affil[1]{Aix Marseille Univ., Univ. de Toulon, CNRS, CPT, UMR 7332, 13288 Marseille, France}
\date{\small\raisebox{0.5em}{Octobre 2026}\vspace{-2.3em}}

\begin{document}

\maketitle
\begin{abstract}
\noindent
In the extended phase space of Donnelly and Freidel, the edge modes of a 
Schwarzschild black hole are dynamical reference frames whose phase 
space is the cotangent bundle of $\mathrm{SL}(2,\mathbb{R})$. 
Gluing the two exterior regions at the bifurcation point leaves a single canonical pair. 
Within the extended phase space, this pair can be described either by Kucha\v{r}'s 
mass and difference of asymptotic times or by the area and the relative rapidity of the two frames, which are two gauge choices of the same system.
The latter is the pair of Carlip and Teitelboim, and 
the relative rapidity is the time shift of the crossed product construction. 
We quantize the frames in this gauge and construct the gravitational 
entangling product of the two exteriors. Requiring the frames to be causal restricts them to the 
semigroup $\mathrm{SL}^+(2,\mathbb{R})$, whose Plancherel measure yields the 
Bekenstein--Hawking entropy as the entanglement entropy between the two 
exteriors, together with a logarithmic correction, both independent of 
the ultraviolet cutoff.
\end{abstract}
\vspace{1.5em}
\begingroup
\small
\setlength{\parskip}{0pt}
\tableofcontents
\endgroup
\clearpage

\section{Introduction}
General relativity is a diffeomorphism-invariant theory. While this statement might seem trivial
today, it was not always so. Einstein himself originally rejected
generally covariant field equations based on the “hole argument” \cite{Norton:1993,Stachel2014}. Consider a region
$H$ of spacetime free of matter and a coordinate transformation $x\mapsto x'$ that
is the identity outside of $H$ but not inside it. The matter distribution is unchanged, while inside
$H$ the transformed metric components $g'_{\mu\nu}(x')$ are different functions than
$g_{\mu\nu}(x)$. By general covariance, the functions $g'_{\mu\nu}$, evaluated at the original
coordinate values, also solve the field equations with the same sources. The same matter
distribution would then determine many different gravitational fields inside $H$, in apparent
contradiction with determinism. Einstein resolved the problem at the end of 1915 by realizing that
the argument assumes that a spacetime point, labelled by its coordinates, has a physical identity
independent of the fields. Instead, the two solutions describe the same physical situation
and ``all our space-time verifications invariably amount to a determination of space-time
coincidences''~\cite{Einstein:1916}. In generally covariant theories, physical observables are therefore relational: given two
fields $A$ and $B$, what one can observe is the value of $A$ at the point where $B$ takes the
value $b$~\cite{Komar:1958,Bergmann:1961,Rovelli:1990ph,Rovelli:2001bz,Dittrich:2005kc,Tambornino:2011vg}.
In other words, observables are only defined with respect to some reference frame.
\par    
One situation in which these considerations play an important role is the definition of local subregions in gravity.
Indeed, observables that one would like to associate with a given subregion may be defined in relation to objects lying outside of it~\cite{Giddings:2005id,Donnelly:2015hta,Goeller:2022rsx}.
Consequently, there is no obvious way to unambiguously define
the algebra of observables associated with a subregion in a
covariant theory~\cite{Casini:2013rba}.
Closely related to this is the fact that the physical phase
space of the theory does not factorize over subsystems.
This complication already arises in gauge theories on a fixed
background.
In electromagnetism, for example, Gauss's law relates the charge
contained in a region to the electric flux through its boundary.
The normal component of the electric field at the boundary is
then shared by the two regions.
Similarly, a closed Wilson loop crossing the boundary can be
split into two open Wilson lines, neither of which is gauge
invariant on its own.
As a result, the full gauge-invariant Hilbert space does not
factorize into a naive tensor product of the Hilbert spaces
associated with each subregion~\cite{Buividovich:2008gq,Donnelly:2011hn,Donnelly:2014gva,Ghosh:2015iwa}.
\par     
In gravitational theories, charges associated with diffeomorphisms also appear at the boundary of the region under consideration~\cite{Regge:1974zd,Wald:1999wa}.
In fact, while the charges associated with symmetries at asymptotic infinity have been an active area of research since the discovery of the BMS group in the sixties~\cite{Bondi1962,Sachs1962,Sachs},
those associated with the boundary of a finite subregion have attracted a lot of interest more recently. In particular, the seminal work of Donnelly and Freidel~\cite{Donnelly:2016auv} (see also~\cite{Freidel:2015gpa,Speranza:2017gxd}) identified the symmetry group associated with such a boundary, now called the corner symmetry group. 
Here, the corner denotes the codimension-2 surface obtained as the intersection of the codimension-1 surface on which the symplectic structure is defined with the boundary of the subregion. These results rely on the covariant phase space formalism, in which the symplectic structure and the charges are constructed directly from the Lagrangian~\cite{GAWEDZKI1972307,Kijowski:1973gi,Crnkovic:1986ex,Crnkovic_1988,KANATCHIKOV199849,Lee:1990nz,Wald:1999wa}.
These corner charges are the gravitational analogue of the normal component of the electric field in electromagnetism, which is the charge of the gauge transformations acting at the boundary. They are the data that the two sides share, and they are responsible for the failure of the factorization of the underlying phase space.
Let us also point out that corner charges are directly related to black hole entropy, since the Wald entropy of a stationary black hole is given by the Noether charge of the horizon Killing vector evaluated at the bifurcation surface~\cite{Wald:1993nt,Iyer:1994ys}.
In the same work, Donnelly and Freidel proposed to resolve this factorization issue by introducing new dynamical fields living on the boundary, which transform under the corner symmetry group, and by gluing subregions through the so-called entangling product. 
These new fields constitute a type of \textit{edge modes}, which had already been investigated in earlier works and have since been studied extensively~\cite{Carlip:1994gy,Balachandran:1994up,Donnelly:2014fua,Donnelly:2015hxa,Geiller:2017xad,Geiller:2017whh,Blommaert:2018rsf,Geiller:2019bti,Giesel:2024xtb}.
It was later understood that they can be seen as dynamical reference frames carrying the information about the complement of the subregion~\cite{Carrozza:2021gju,Carrozza:2022xut}.
This construction was carried out with some details for Yang--Mills theory in \cite{Donnelly:2016auv}. The complete treatment of the gravitational case however, would require the full representation theory of the corner symmetry group $\mathrm{Diff}(S)\ltimes\mathrm{SL}(2,\mathbb{R})^S$, as well as some knowledge about the harmonic analysis on it.
While this is not yet accessible, some work was done in this direction by using the corner symmetry group as an organizing principle to describe the classical or quantum gravitational physics associated with a subregion
This is known in the literature as the local holography program or corner proposal~\cite{Freidel:2020xyx,Freidel:2020svx,Freidel:2020ayo,Donnelly:2020xgu,Freidel:2021cjp,Ciambelli:2021vnn,Ciambelli:2021nmv,Ciambelli:2022cfr,Ciambelli:2022vot,Freidel:2023bnj,Ciambelli:2023bmn,Ciambelli:2024swv,Ciambelli:2024qgi,Varrin:2024sxe,Ciambelli:2025ztm,Neri:2025fsh,Kowalski-Glikman:2025arealaw,Varrin:2025okc,Neri:2026vak,Varrin:2026bmb}.
\par 
In the edge mode approach, the entropy area term appears as an entanglement 
edge term associated with a natural measure on the gauge group
in analogy with the edge mode contribution in gauge theory~\cite{Donnelly:2014fua,Donnelly:2014gva}. 
In two dimensions, this edge term reproduces the Bekenstein--Hawking
entropy of Jackiw--Teitelboim gravity, formulated as an
$\mathrm{SL}(2,\mathbb{R})$ BF theory~\cite{Lin:2017uzr,Lin:2018xkj}. In three dimensions, the entropy of the BTZ black hole is recovered as the
entanglement entropy of gravitational edge modes, once the Chern--Simons
description is refined to a topological field theory based on the quantum
semigroup $\mathrm{SL}^+_q(2,\mathbb{R})$~\cite{Mertens:2022ujr},
with the corresponding Plancherel contribution to the entropy also identified
in a derivation from boundary CFT data~\cite{Bao:2025rt}.
In another related approach, the reference frames are treated quantum mechanically directly within the theory itself.
This is a well motivated perspective, as nature is fundamentally quantum and the reference frames should be treated as any other physical system.
This idea has its origin in two independent lines of work from the sixties. In quantum foundations, 
Aharonov and Susskind argued that superselection rules, such as the one forbidding 
superpositions of states with different charges, can be circumvented when the reference system 
is itself treated as a quantum system~\cite{Aharonov:1967zza,Aharonov:1967zz}. 
For closed universes, DeWitt realized that the quantum state of the gravitational field does not 
depend on an external time, so that evolution has to be described relative to physical degrees 
of freedom of the system itself~\cite{dewitt1967quantum}, an idea later developed into relational 
quantum dynamics~\cite{Page:1983uc,Rovelli:1990pi}. 
Aharonov and Kaufherr then showed that quantum mechanics can be consistently formulated 
relative to a quantum reference frame of finite mass, without appealing to an idealized 
classical frame~\cite{AharonovKaufherr1984}. These ideas were further developed in quantum 
information theory, where reference frames are seen as physical resources that may or may not 
be shared between 
parties~\cite{Bartlett:2006tzx,Spekkens:2008acy,Angelo:2011skm,Palmer:2013zza,Smith:2016fmm,Loveridge:2016tnh}. 
More recently, the study of relational quantum dynamics has led to the modern concept of 
quantum reference frames~\cite{Giacomini:2017zju,Vanrietvelde:2018pgb,Hohn:2018toe,delaHamette:2020dyi,Hoehn:2019fsy,delaHamette:2021oex,Carette:2023wpz,Hoehn:2023ehz,Kabel:2023jve,Kabel:2024lzr,Geng:2024dbl}.
\par   
In gravity, quantum reference frames, and the closely related crossed product \cite{Klinger:2023qna}, have recently been used to define the algebra of observables of a subregion and its 
entropy~\cite{Witten:2021unn,Chandrasekaran:2022cip,Chandrasekaran:2022eqq,Jensen:2023yxy,DeVuyst:2024khu,DeVuyst:2024fxc,Geng:2025bcb,Chen:2025tbh}.
Boost edge modes obtained by splitting the bulk Hilbert space across a horizon, together with the associated crossed product algebra, have also recently appeared in Jackiw--Teitelboim gravity in the context of the firewall paradox~\cite{Chandrasekaran:2026gvk}.
In this approach the entropy area term appears from a different perspective. The algebra of the subregion can be extended by a degree of
freedom that acts as a clock for the boost-like flow generated by the
horizon Killing vector, or more generally for the modular flow.
The resulting algebra of observables is of type 
II, and its entropy agrees with the generalized entropy up to a state-independent 
constant. The area term then arises in the limit 
$G_N\to0$, either from the relation between the ADM mass and the horizon area
or from the first law of local subregions.
\par    
The two approaches are thus close in spirit, since both complete the
subregion by a boundary degree of freedom, but they differ in how the
area term arises. In the edge mode approach it comes from the measure on
the symmetry group, which has so far been supplied by a reformulation of
the gravitational theory as a topological gauge theory of a Lie group, available only in two and
three dimensions. In the quantum reference frame approach it comes from the constraint
relating the energy of the clock to the area, in the limit $G_N\to0$, and
the symmetry group plays no role. Although the crossed product has been shown to arise from the quantization of an extended phase space~\cite{Klinger:2023qna}, the two mechanisms producing the area term have not been related in full generality. In particular, it is not known whether the edge mode measure can produce the area term without a gauge theory formulation, nor whether it is possible to realize the clock of quantum reference frames as a Donnelly--Freidel type edge mode.
In this work, we make this connection explicit for static spherically symmetric black holes, without relying on a gauge theory formulation. 
The dynamical reference frame does not come with an ad hoc structure. Once gauge invariance of the subregion phase space is imposed, its phase space, symmetry group and charges are fixed by the Donnelly--Freidel symplectic structure of the dilaton theory obtained by reducing four-dimensional Einstein--Hilbert gravity on spherically symmetric configurations. The remainder of this introduction describes this construction and its main results.
\par   
When considering a subregion of spacetime,
some diffeomorphisms acquire a physical charge on the boundary of that subregion, breaking gauge invariance. In order to restore it, one must add edge modes in the
form of a frame, living on the boundary. Restricting to diffeomorphisms that do not move the position of the corner, this new degree of freedom is associated with the symplectic
structure $\mathrm{T}^*\spl{2}$ through the Donnelly--Freidel prescription.
When the corner is seen as the entangling surface between two subregions, there is no boundary and the individual orientation of each frame is not physical.
Starting from $\mathrm{T}^*\spl{2}\times\mathrm{T}^*\spl{2}$, one copy for the frame 
of each subregion, the entangling product imposes the matching of charges and 
quotients by the diagonal relabelling of the two frames. In gravity, each region also carries a Gauss law, 
which identifies the left momentum of its frame with the corner charge of its fields. The glued spacetime 
has a single smooth geometry at the corner, so that the left momenta of the two frames must agree. 
Only the diagonal action of spacetime diffeomorphisms then remains gauge, together with the boost that 
preserves the corner geometry. At the bifurcation point, this boost is generated by the Killing vector. 
It shifts the relative boost of the two frames together with the relative Killing time of the two 
exteriors, and fixing the latter leaves the former as the physical variable. 
What remains after this gluing and gauge fixing is the relative rapidity between the frames, which is conjugate to the area.
\par   
Interestingly, the procedure therefore recovers the canonical pair of Carlip and
Teitelboim~\cite{Carlip:1993sa}, the relative rapidity of the two frames
playing the role of the Euclidean opening angle at the bifurcation surface.
In a sense, this could have been anticipated. By Birkhoff's theorem, spherically symmetric vacuum gravity has no local 
degrees of freedom, and Kucha\v{r} showed that its reduced phase space is two-dimensional, spanned by the mass and the difference of the Killing times at the two asymptotic ends~\cite{Kuchar:1994zk}. 
A consistent gluing must therefore leave exactly one canonical pair at the corner. The non-trivial statement is that it does so, 
and that it locates this pair at the bifurcation point as the relative orientation of 
the two edge frames. 
As already noted by Carlip and Teitelboim, and made explicit in the canonical analyses of~\cite{Louko:1994tv,Kunstatter:1997my}, this pair is canonically equivalent, through the first law, to Kucha\v{r}'s. 
In the extended phase space, the two are in fact related by a gauge transformation, a boost acting on one side of the corner only.
The relative rapidity is in turn the timeshift of the two-sided crossed product construction~\cite{Chandrasekaran:2022eqq}.
Edge modes, the Carlip--Teitelboim pair, Kucha\v{r}'s time and the crossed product 
clock thus appear as different facets of a single relative boost between the two edge frames.

\par   
Finally, we require the frames to be causal, i.e.\ to map future-directed directions of the reference plane to future-directed directions of spacetime. This restricts the frames to the semigroup $\mathrm{SL}^+(2,\mathbb{R})$ of 
matrices with positive entries.
The entanglement entropy of the glued state is then controlled by the 
Plancherel measure of this semigroup, $\dd\mu(k)\propto k\sinh(2\pi k)\,\dd k$, where $k$ labels the representation and 
is fixed by the area. This measure grows exponentially, while the 
Plancherel measure of the group $\mathrm{SL}(2,\mathbb{R})$ only grows linearly and would yield a logarithm of the area. 
The gauge constraints described above (Gauss constraint, frame matching and continuity) maximally entangle both factor of the frame wavefunction across the corner, which each contribute one power of the Plancherel density. Together they give exactly the Bekenstein--Hawking entropy
\begin{equation*}
    S_{\mathrm{edge}}=\frac{A}{4G_N\hbar}+2\ln\frac{A}{4G_N\hbar}+\dots,
\end{equation*}
where the dots denote state-dependent corrections controlled by the relative width of the area distribution, and non-perturbative corrections. All of these terms are independent of the ultraviolet cutoff, which only enters a state-independent constant.\par    
The paper is organized as follows. In Section~\ref{sec:dimensionallyreducedblackhole} we describe the symplectic structure of the dimensionally reduced black hole. We introduce the
extended phase space of \cite{Donnelly:2016auv} and compute the Noether charges associated with the corner preserving symmetries. This allows us to introduce the concept of frame and to identify the
algebra of charges to be the one of $\spl{2}$. We then specialize to Schwarzschild solution in Kruskal--Szekeres coordinates with the corner situated at the bifurcation point, which enables us
to identify the special linear group's Casimir with the area of the entangling surface. Next, in Section~\ref{sec:quantumandreferenceframe},
we introduce the notion of quantum reference frame associated with the corner symmetries. At the same time, we also discuss the classical version of the reference frame to gather intuition and to identify
the dynamical pair as the one of Carlip and Teitelboim and relate it to Kucha\v{r}’s time and to the clock of the crossed product construction. We end the section by constructing the entangling product of \cite{Donnelly:2016auv}.
Finally, we discuss the entanglement entropy of the resulting glued state and the emergence of the area term in Section~\ref{sec:entanglemententropy}. Conclusions and comparison to other works
are given in Section~\ref{sec:conclusion}. The gauge constraint analysis of the gluing and the mathematical details about the semigroup $\mathrm{SL}^+(2,\mathbb{R})$ are respectively relegated to Appendix~\ref{app:constraints} and Appendix~\ref{app:SLplus}.

\section{Dimensionally Reduced Black Hole}\label{sec:dimensionallyreducedblackhole}
In order to describe the quantum mechanics of the edge modes on the black hole horizon, we start by describing the gravitational
system as a two-dimensional dilaton theory. We then introduce the extended symplectic structure of \cite{Donnelly:2016auv} by extracting
a symplectic potential from the dimensionally reduced action and by introducing a dynamical coordinate system. The latter allows us to add a corner term to the symplectic potential which renders it diffeomorphism invariant.
We then compute the Noether charges associated with the corner preserving transformation and specialize to the Schwarzschild solution in the Kruskal--Szekeres gauge. Throughout this section, we use the covariant phase space formalism.
We refer the unfamiliar reader to see \cite{Speziale:2025lkm} for a modern review.
\par 
We start with the Einstein--Hilbert action in a four-dimensional spacetime
\begin{equation}\label{eq:action}
    S_{\mathrm{EH}} = \frac{1}{16\pi G_N}\int_{M}\dd^4x\,\sqrt{-g}\, R_4 .
\end{equation}
Consider a spherically symmetric metric
\begin{equation}
    \dd s^2 = g_{ij}\dd x^i \dd x^j + \rho(x^i)^2\dd \Omega_S^2,
\end{equation}
where $i,j=0,1$ and $\dd\Omega_S^2$ is the metric of the round two-sphere. The theory studied in this paper is obtained by plugging the above
form of the metric inside the action \eqref{eq:action},
integrating over the sphere
and discarding total derivatives.\footnote{While the integration over the sphere produces a boundary term, it vanishes on null hypersurfaces and can thus be discarded in the present consideration.} The reduced action is then given by
\begin{equation}\label{eq:2Daction}
    S_{\mathrm{EH}} = \int_{M_2}\dd^2 x \,\sqrt{-\hg}\qty(\Phi\hat R_{2} + \frac{2}{\ell^2}\,\Phi^{-\frac12}),
\end{equation}
where $\ell$ is an arbitrary length scale that drops out of all physical results and we introduced the dilaton and conformally rescaled metric
\begin{equation}
    \Phi = \frac{\rho^2}{4G_N},
    \qquad
    \hg_{ij} = \frac{\rho \ell^2}{\qty(4 G_N)^{\frac32}}\,g_{ij}.
\end{equation}
Varying $S_{\mathrm{EH}}$ gives, up to terms proportional to the equations of motion, the total derivative
$\partial_i\theta^i$ with symplectic potential current
\begin{equation}\label{eq:symplecticpotentialcurrent}
    \theta^i= \sqrt{-\hg}\qty(\Phi\, \hat \nabla_j \theta^{ij} - \theta^{ij}\, \hat \nabla_j \Phi),
    \qquad
    \theta_{ij} = \delta \hg_{ij} - \hg_{ij}\hg^{kl}\delta \hg_{kl} .
\end{equation}
\par
In order to describe the symplectic structure associated with subsystems, we consider a Cauchy slice $\Sigma$ and split it into two complementary pieces $\Sigma_{\rm I}$ and $\Sigma_{\rm II}$. The shared boundary $S$ is a codimension-2 surface that we call the corner.
The subregions $\rm I$ and $\rm II$ are the domains of dependence of $\Sigma_{\rm I}$ and $\Sigma_{\rm II}$, and are bounded by the null hypersurface $\mathcal{N}$ of constant dilaton that passes through $S$.\footnote{In the two-dimensional theory, $\mathcal{N}$ is a null line and $S$ a point. From the point of view of the original four-dimensional theory, they are a null hypersurface and a two-sphere.}
Denoting the subregions collectively by $\mathcal{R} \in \qty{\rm I, \rm II}$, their presymplectic potential is obtained by integrating the current \eqref{eq:symplecticpotentialcurrent} over the respective Cauchy slices
\begin{equation}\label{eq:symplecticpotential}
    \Theta_{\mathcal{R}} = \int_{\Sigma_{\mathcal{R}}} \dd\Sigma_i\,\theta^i ,
    \qquad
    \dd\Sigma_i=\epsilon_{ij}\,\dd x^j\eval_{\Sigma},
\end{equation}
where $\epsilon_{ij}$ is the Levi-Civita symbol such that $\epsilon_{01}=1$ in coordinates where $x^0$ increases towards the future. Note that the orientation of $\Sigma$ induces opposite orientations of $S$ as a boundary of $\Sigma_{\rm I}$ or $\Sigma_{\rm II}$. The presymplectic form is then given by the field-space derivative of the presymplectic potential
\begin{equation}
    \Omega_{\mathcal{R}} = \delta \Theta_{\mathcal{R}}.
\end{equation} 
\par 

Let us now introduce the extended phase space of Donnelly and Freidel~\cite{Donnelly:2016auv}. 
Diffeomorphisms are gauge redundancies of the theory, and should therefore be degenerate directions of the presymplectic form. Contracting $\Omega_{\mathcal R}$ with the field-space vector field $V_\xi$ generating a diffeomorphism that preserves the corner, $\xi\eval_S=0$, one finds however
\begin{equation}\label{eq:Iomega}
    -I_{V_\xi}\Omega_{\mathcal R}\,\hat{=}\,\delta\qty(\sigma^{\mathcal R}_S\,Q_\xi\eval_S),
\end{equation}
where the hat above the equal sign denotes equality on shell, $\sigma_S^{\mathcal{R}}$ is defined below and
\begin{equation}\label{eq:Qxi}
    Q_\xi=\sqrt{-\hg}\,\epsilon_{ij}\qty(\Phi\,\hat\nabla^j\xi^i-2\,\xi^i\,\hat\nabla^j\Phi).
\end{equation}
This quantity does not vanish when $\xi$ has support near or at the corner. 
In the presence of a boundary, such diffeomorphisms are thus not gauge but carry non-vanishing charges. 
The situation is similar to the one in Chern--Simons theory, where gauge 
transformations that do not vanish at the boundary change the action by a boundary term.
There, gauge invariance is restored by boundary degrees of freedom, described by a chiral Wess--Zumino--Witten 
model~\cite{Witten:1988hf,Elitzur:1989nr}, which are the edge modes responsible for the edge currents of the 
quantum Hall effect~\cite{Halperin:1981ug,Wen:1990se,Wen:1992vi,Balachandran:1991dw}. 
Similarly, following~\cite{Donnelly:2016auv},
gauge invariance is restored here by introducing a coordinate system for each subregion, $X_{\mathcal R}:U_r^{\mathcal R}\to M$. Here $U_r^{\mathcal R}$ is an open subset of a reference space $M_r$, which can be thought of as the space of labels that a physical reference frame, e.g.\ the clocks and rods of an observer, assigns to events. Furthermore, we assume that $U_r^\mathcal{R}$ is a neighborhood of the reference slice $\Sigma_r^{\mathcal R}$, which is mapped to $\Sigma_{\mathcal R}$. The reference corner is then defined as the intersection $S_r=\Sigma_r^{\rm I}\cap\Sigma_r^{\rm II}$. Both coordinate systems map the reference corner to the physical one, $X_{\rm I}(S_r)=X_{\rm II}(S_r)=S$, but are otherwise independent; in particular, their first derivatives at $S_r$ need not agree. In what follows, we will sometimes drop the label $\mathcal{R}$ on $X$ and related quantities when no confusion can arise.
The symplectic form can then be made
invariant under diffeomorphisms by adding a corner term to the symplectic potential built from the variations of $X$.
Since we are considering field space variations of the coordinate system, it is itself added to the phase space. This is the origin of the dynamical reference frame that we will consider later. 
The extended symplectic form is then defined on the extended phase space \cite{Donnelly:2016auv,Speranza:2017gxd}
\begin{equation}\label{eq:DFsymplectic}
  \ODF\qty[\hat{g},\Phi,X] = \Omega_{\mathcal{R}}\qty[\hat{g},\Phi]+ \int_{S_r}X^*\qty(\iota_\chi \theta + \frac12 \iota_\chi \iota_\chi L + \delta Q_\chi + \ldv{\chi}Q_\chi),
\end{equation}
where we defined the vector field $\chi = \delta X \circ X^{-1}$. Diffeomorphism invariance can be checked explicitly by using the action of diffeomorphisms on the coordinate system $I_{V_\xi}\chi=-\xi$.
We will sometimes call the coordinate system edge modes in analogy with the Chern--Simons case.
Because of the dimensional reduction, the corner is simply a point. The integral of a $0$-form over $S_r$, oriented as the boundary of the reference slice of the region $\mathcal R\in\{\rm I, \rm II\}$, is therefore its value at $S=X(S_r)$ times an orientation sign $\sigma^{\mathcal R}_S=\pm1$, with $\sigma^{\rm I}_S=-\sigma^{\rm II}_S$. Writing $\theta=\epsilon_{ij}\theta^i\dd x^j$, the symplectic form of the region $\mathcal R$ becomes
\begin{equation}\label{eq:odfin2d}
    \Omega^{\rm DF}_{\mathcal R}=\Omega_{\mathcal R}+\sigma^{\mathcal R}_S\qty[\epsilon_{ij}\theta^i\chi^j+\mathcal{L}\,\chi^0\chi^1+\delta Q_\chi+\chi^i\partial_iQ_\chi]\eval_{S},
\end{equation}
where $\mathcal{L}$ is the bulk Lagrangian density of the dilaton theory.
\par
The symplectic form \eqref{eq:odfin2d} is invariant under all spacetime diffeomorphisms, including those acting at the corner: $I_{V_\xi}\Omega^{\rm DF}_{\mathcal R}\,\hat{=}\,0$ for every $\xi$ \cite{Donnelly:2016auv,Speranza:2017gxd}. Spacetime diffeomorphisms are therefore pure gauge. As shown in Section~\ref{sec:quantumandreferenceframe}, the charge \eqref{eq:Iomega} of their action on the fields is cancelled by the charge of their action on the coordinate system. 
The physical transformations are instead the surface symmetries, which relabel the reference space while leaving the dynamical fields untouched. In the language of dynamical reference frames, these are the frame reorientations of \cite{Carrozza:2021gju,Carrozza:2022xut}. To a vector field $w$ on $U_r$ we associate the field-space vector field $R_w$ defined by
\begin{equation}
    I_{R_w}\delta\varphi=0,
    \qquad
    I_{R_w}\chi=W,
    \qquad
    W\equiv X_*w,
\end{equation}
which implements $X\to X\circ z$, with $z$ the flow of $w$ inside the reference space. In the above equation and in the remainder of this paper, $\varphi$ collectively denotes all the fields of the theory.
Contracting it into the symplectic form gives \cite{Speranza:2022lxr}
\begin{equation}\label{eq:obstruction}
    -I_{R_w}\Omega^{\rm DF}_{\mathcal R}\,\hat{=}\,\delta\qty(\sigma^{\mathcal R}_S\,Q_W\eval_S)-\sigma^{\mathcal R}_S\qty[Q_{\delta W+[\chi,W]}-\iota_W\qty(\theta+\iota_\chi L+\dd Q_\chi)]\eval_S .
\end{equation}
For a field-independent $w$ one has $\delta W+[\chi,W]=X_*\delta w=0$, and the first obstruction term vanishes. The second vanishes when $w$ preserves the corner, which in two dimensions means $w|_{S_r}=0$. These surface symmetries are Hamiltonian vector fields of $\ODF$ with
\begin{equation}\label{eq:Hw}
    H^{\mathcal R}_w=\sigma^{\mathcal R}_S\,Q_W\eval_S=\sigma^{\mathcal R}_S\sqrt{-\hg}\,\epsilon_{ij}\,\Phi\,\hat\nabla^jW^i\eval_S ,
\end{equation}
where the second term of \eqref{eq:Qxi} drops because $W|_S=0$. Surface symmetries that move the corner, $w|_{S_r}\neq0$, are non-hamiltonian because of the last term of \eqref{eq:obstruction}.
\par
In order to understand which algebra the above charges realize, we express 
the charges in terms of the jets of the fields at the corner. 
Let $x^i$ be coordinates on $M$ adapted to the corner $S= \qty{x^i = 0}$ and $y^a$ be coordinates on $U_r$ adapted to the corner $S_r=\{y^a=0\}$. The dynamical coordinate system defines the edge frame
\begin{equation}\label{eq:edgeframe}
    (e_{\mathcal{R}})^i{}_a=\partial_aX_{\mathcal{R}}^i\eval_{S_r},
\end{equation}
which maps the reference normal plane to the spacetime one. The name \textit{frame} refers to the fact that these fields can be interpreted as dynamical reference frames as we will see in Section~\ref{sec:quantumandreferenceframe}.
A corner-preserving surface symmetry has the expansion $w=w^a_{(1)b}\,y^b\partial_a+\mathcal O(y^2)$. Since $w$ vanishes at $S_r$, the first jet of $W=X_*w$ at $S$ is $\partial_jW^i|_S=e^i{}_a\,w^a_{(1)b}\,(e^{-1})^b{}_j$. Inserting it into \eqref{eq:Hw} gives
\begin{equation}\label{eq:charges}
    H^{\mathcal R}_w=\sigma^{\mathcal R}_S\,(N_r)_a{}^b\,w^a_{(1)b},
    \qquad
    (N_r)_a{}^b=(e^{-1})^b{}_j\,N_i{}^j\,e^i{}_a,
\end{equation}
where
\begin{equation}\label{eq:chargesNij}
   N_i{}^j=\qty(\Phi\,\epsilon^{jl}\,\frac{\hg_{il}}{\sqrt{-\hg}})\eval_{S}.
\end{equation}
From equation \eqref{eq:charges} we see that $N_r$ and $N$ have the same trace and determinant. Since $N_i{}^i=0$, only the traceless part of $w_{(1)}$ is charged. For field-independent surface symmetries, the charges represent the algebra of vector fields, $\{H^{\mathcal R}_v,H^{\mathcal R}_w\}=-H^{\mathcal R}_{[v,w]}$. Restricting to representatives with vanishing second jet, this yields
\begin{equation}\label{eq:Nrbracket}
    \{(N_r)_a{}^b,(N_r)_c{}^d\}=\sigma^{\mathcal R}_S\qty(\delta^b{}_c\,(N_r)_a{}^d-\delta^d{}_a\,(N_r)_c{}^b),
\end{equation}
which is $\mathfrak{sl}(2,\mathbb R)$. Although the orientation sign appears on the right hand side, not that  the variables $\sigma^{\mathcal R}_S N_r$ still satisfy the same algebra on both sides of the corner.
There is an important distinction between the matrices $N$ and $N_r$.
While the matrix $N$ depends only on the dilaton and metric at the corner, the charges associated with a particular diffeomorphism do not measure this matrix itself. Instead, it measures its components
$N_r$ in the basis defined by the frame. For a fixed reference generator $w$, the charge $H^{\mathcal{R}}_w$ therefore also depends on the particular frame in which one expresses it.
In other words, it is a relational observable, describing the corner geometry as seen by the frame.
\par 
We now consider the family of static spherically symmetric solutions
\begin{equation}\label{eq:sssmetric}
    \dd s^2 = -f(\rho)\dd t^2 + \frac{\dd \rho^2}{f(\rho)}+ \rho^2 \dd\Omega_S^2,
\end{equation}
where the largest root $\rho_h$ of $f$ is assumed to be simple and defines the horizon with surface gravity $\kappa = \frac12 f'(\rho_h) \neq 0$.
For the action \eqref{eq:action}, Birkhoff's theorem implies that $f(\rho)=1-2G_NM/\rho$, i.e.\ the solution is Schwarzschild. We nevertheless keep $f$ general, since the construction below only uses the existence of a simple horizon with $\kappa\neq0$. It therefore applies, for instance, to Schwarzschild--(A)dS once a cosmological constant is included in \eqref{eq:action}.
Introducing the tortoise coordinate $\rho_*=\int^\rho \dd\rho'/f(\rho')$, with the integration constant fixed such that $\rho_*\simeq\frac{1}{2\kappa}\ln|\rho-\rho_h|$ near the horizon, the null coordinates $u=t-\rho_*$, $v=t+\rho_*$ and the Kruskal-like coordinates $U=-e^{-\kappa u}$, $V=e^{\kappa v}$, the metric becomes
\begin{equation}
    \dd s^2 = \frac{f(\rho)}{\kappa^2 U V} \dd U \dd V + \rho^2 \dd\Omega_S^2,
    \qquad UV=-e^{2\kappa\rho_*}.
\end{equation}
The maximal extension is obtained by allowing both signs of $U$ and $V$, within the range where $\rho(UV)$ is defined. The future and past horizons are $U=0$ and $V=0$, and the bifurcation point is $U=V=0$. In Kruskal--Szekeres coordinates $U=T-R$, $V=T+R$,
the metric is conformally flat
\begin{equation}
    \dd s^2 = \frac{f(\rho)}{\kappa^2}e^{-2\kappa \rho_*}\qty(-\dd T^2 + \dd R^2) + \rho^2 \dd\Omega_S^2,
\end{equation}
with a conformal factor that depends on $T$ and $R$ only through $\rho(T^2-R^2)$.
\par
\begin{figure}[H]
    \centering
    \includegraphics[width=.6\textwidth]{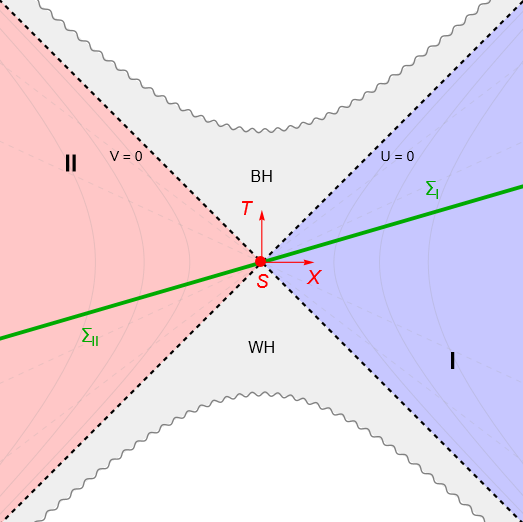}
    \caption{Kruskal--Szekeres diagram of the maximally extended black hole. The exterior regions $\rm I$ and $\rm II$ are separated by the horizons $U=0$ and $V=0$, which meet at the bifurcation point $S$. The Cauchy slice $\Sigma=\{T=R\tanh(\kappa t_0)\}$ splits into $\Sigma_{\rm I}$ and $\Sigma_{\rm II}$ at the corner $S$. Wiggly lines denote the singularities, dashed light grey lines denote surfaces of constant Schwarzschild time $t$ and light gray hyperbolas are surfaces of constant $\rho$.}
    \label{fig:blackhole}
\end{figure}

We choose the Cauchy slice $\Sigma=\{T=R\tanh(\kappa t_0)\}$, oriented along increasing $R$. We also choose both coordinate systems such that the reference corner sits at the bifurcation point: $X_{\rm I}(S_r)=X_{\rm II}(S_r)=\{T=R=0\}$. Note that, since the coordinate system is part of the field space, this is part of the field configuration.
Next we consider the subregions $\rm II$ and $\rm I$ as the left ($R<0$) and right ($R>0$) exterior regions, so that $\sigma^{\rm I}_S=-1$ and $\sigma^{\rm II}_S=+1$.
The maximally extended spacetime and the chosen Cauchy slice are depicted in Figure \ref{fig:blackhole}. 
The dilaton and the conformally rescaled metric read
\begin{equation}
    \Phi=\frac{\rho^2}{4G_N },
    \qquad
    \hg_{ij}\dd x^i\dd x^j=e^{2\hat\sigma(\rho)}\qty(-\dd T^2+\dd R^2),
\end{equation}
where
\begin{equation}
    e^{2\hat\sigma}=\frac{\rho \ell^2}{\qty(4G_N )^{\frac32}}\,\frac{f(\rho)}{\kappa^2}\,e^{-2\kappa\rho_*}.
\end{equation}
For this value of the metric, it is easy to see that the charges \eqref{eq:chargesNij} at the bifurcation corner are given by
\begin{equation}\label{eq:Ncharges}
    N_T{}^R=N_R{}^T=\Phi_h,
    \qquad
    N_T{}^T=N_R{}^R=0,
    \qquad
    \Phi_h=\frac{\rho_h^2}{4G_N }.
\end{equation}
\noindent
Let us now comment on diffeomorphisms that move the boundary. For a vector field that does not vanish at the corner, the Noether potential also contains a term linear in its value there
\begin{equation}
    Q_\xi\eval_S=t_i\,\xi^i_{(0)}+N_i{}^j\,\xi^i_{(1)j},
    \qquad
    t_i=\sqrt{-\hg_{(0)}}\;\epsilon_{kl}\,\hg_{(0)}^{lm}\qty(\Phi_{(0)}\,\hat\Gamma^k{}_{mi}-2\,\delta^k{}_i\,\Phi_{(1)m}),
\end{equation}
with $\hat\Gamma^k{}_{mi}$ built from $\hg_{(0)}$ and its first derivative $\hg_{(1)}$ at the corner, and $\Phi_{(1)}$ the
first derivative of the dilaton there. In the DF phase space these would-be translation charges are not Hamiltonian, because of 
the last term of \eqref{eq:obstruction}. In the extended phase space of \cite{Ciambelli:2021nmv}, where they are Hamiltonian, the translation charges $t_i$ together with $N_i{}^j$ form the algebra of the (two dimensional) extended corner symmetry group
\begin{equation}
    \mathrm{ECS}_2 = \spl{2} \ltimes \mathbb{R}^2.
\end{equation}
Within the CLP formalism \cite{Ciambelli:2021nmv}, the Noether charges provide momentum maps that land in the coadjoint orbits of the full ECS group. The coadjoint orbits of the $\mathrm{ECS}_2$ are discussed in detail in \cite{Ciambelli:2022cfr,Neri:2025fsh}.
For the situation at hand however, the DF description we used above is actually justified. 
To see why, recall that the coadjoint orbits of a semidirect product $G = H\ltimes A$, with $A$ abelian, are organized by the orbits of $H$ acting on the dual $A^*$.
In the familiar case of Poincaré symmetry, these are the momentum hyperboloids classified by the mass invariant $m^2 = -p_\mu p^\mu$. By analogy, the orbits of $H$ acting on the dual $A^*$ are often called momentum orbits.
In the case of the $\mathrm{ECS}_2$, the action of $\spl{2}$ is 
transitive on the punctured plane $\mathbb{R}^2_\circ = \mathbb{R}^2 \setminus \qty{(0,0)}$. There are therefore two momentum orbits: the punctured plane and the origin.
For the solution at hand, $\Phi$ and $\hat\sigma$ depend on $T$ and $R$ only through $T^2-R^2$, so their first derivatives vanish at the bifurcation point and the would-be translation charges vanish. This corresponds to the momentum orbit being the origin.
In the next step, we look for the little group $H_{p_*} \subset H$, that stabilizes a representative of a coadjoint orbit $p_*\in A^*$.
For the origin, it is clear that this little group is the whole of $\spl{2}$. The coadjoint orbits of vanishing momentum are therefore isomorphic to the
coadjoint orbits of $\spl{2}$. Looking only at the corner-preserving diffeomorphisms is now justified by this fact.
 The coadjoint orbits of $\spl{2}$ are
classified by the quadratic Casimir, which for the charge matrix $N_i{}^j$
is simply its determinant, $\mathcal{C}=\det N$. This quantity is invariant
under the coadjoint action and insensitive to the orientation sign $\sigma^{\mathcal{R}}_S$. Orbits with $\mathcal{C}>0$ are elliptic (one sheet of a two-sheeted hyperboloid) and are quantized by the discrete series,
orbits with $\mathcal{C}<0$ are hyperbolic (one-sheeted hyperboloids) and
are quantized by the principal series, and $\mathcal{C}=0$ corresponds to
the nilpotent cones and the origin.\par 

In order to identify which orbit describes our situation,
we choose both coordinate systems such that the reference coordinates agree with $(T,R)$ to first order at the corner, $e_{\rm I}=e_{\rm II}=\mathds{1}$, so that $N_r=N$ on both sides. This is the choice of a representative point in the extended phase space, not a gauge fixing of the relative orientation of the two frames, which is a dynamical variable discussed in Section~\ref{sec:quantumandreferenceframe}. Taking the fields in Kruskal form on the straight slice $\Sigma$ is, on the other hand, a gauge choice, as shown there. Choosing instead $e_{\rm I}$ to differ from $e_{\rm II}$ by a boost of the $(T,R)$ plane would leave the charges below unchanged, since such a boost commutes with $N$ in \eqref{eq:Ncharges}.\footnote{More generally, for a frame $e_{\mathcal R}=h\in\spl{2}$, the charges $N_r^{\mathcal R}=h^{-1}Nh$ move along the coadjoint orbit of $N$. The point reached by the moment map changes, but the orbit, and hence the discussion of this section, does not.}
\begin{equation}\label{eq:noethercharges}
    H^{\mathcal R}_w=\sigma^{\mathcal R}_S\,\Phi_h\qty(w^0_{(1)1}+w^1_{(1)0}).
\end{equation}
Among the $\mathfrak{sl}(2,\mathbb R)$ generators $w_1=\frac12(y^1\partial_0+y^0\partial_1)$, $w_2=\frac12(y^1\partial_0-y^0\partial_1)$ and $w_3=\frac12(y^1\partial_1-y^0\partial_0)$, only $w_1$ is charged. Its image $W_1=X_*w_1$ agrees at first order at the corner with the boost $\frac12(R\partial_T+T\partial_R)=\frac12(V\partial_V-U\partial_U)=\frac1{2\kappa}\partial_t$. We can finally write the
charges for the solution \eqref{eq:sssmetric} and the coordinate system placing the corner at the bifurcation point
\begin{equation}
    H^{II}_{w_1}=-H^{I}_{w_1}=\Phi_h=\frac{\rho_h^2}{4G_N},
    \qquad
    H_{w_2}=H_{w_3}=0.
\end{equation}
We note that this is the standard result of Wald that the area is proportional to the Noether charge associated with the boost \cite{Wald:1993nt}.
The Noether charges \eqref{eq:noethercharges} provide a moment map
from a given field configuration to a point on the coalgebra $\mathfrak{sl}(2,\mathbb{R})^*$. We can then compute the value of the Casimir on the orbit passing through that point.
At the bifurcation point, we can use equation \eqref{eq:noethercharges} to compute
\begin{equation}\label{eq:casimirvalueorbits}
    \mathcal{C}_{\mathrm{cl}}=\det N=-\Phi_h^2<0 .
\end{equation}
The corner therefore lies on a hyperbolic coadjoint orbit. Its quantization is the principal series
representation with label $k=\Phi_h$, in which the quadratic Casimir
operator takes the value $-\qty(\frac14+k^2)$. The shift $-\frac14$ is
the usual ordering correction between the classical orbit invariant and the Casimir operator.
If we reinstate $\hbar$ for a moment, we observe that the asymptotic regime $k\gg 1$ corresponds to
\begin{equation}
    \rho_h^2 \gg 4 \ell_P^2.
\end{equation}
In other words, it is the classical limit where the Planck area is negligible with respect to the size of the black hole.
\section{Dynamical Reference Frames}\label{sec:quantumandreferenceframe}
We now turn to the discussion of dynamical reference frames. Here, dynamical means that the frames are
part of the phase-space and they have a conjugate momentum. In order to better understand the kinematics of these frames, 
we will start by revisiting the corner term in the DF symplectic potential \eqref{eq:DFsymplectic}. We then move on to the quantum theory where we construct the entangling product and identify
the remaining degree of freedom as the relative rapidity between the two frames. This result is then studied in the classical setting where we find that the symplectic form reduces to the one of Carlip and Teitelboim after gluing.
\par   
In order to understand the phase space associated with the edge mode $X$ and the frame $e= \dd X\eval_{S_r}$, let us look at the symplectic structure again.
Since we only consider diffeomorphisms that do not move the corner, we impose the condition $\delta X\eval_{S_r}=0$
which implies $\chi\eval_{S} = 0$. The corner contribution to the symplectic form \eqref{eq:odfin2d} can then be written
\begin{Align}
    \Omega^{X}_{\mathcal{R}} &= \sigma_S^{\mathcal{R}}\delta Q_\chi\eval_S\\
     &= \sigma_{S}^{\mathcal{R}} \qty(\delta N\downup{i}{k} (\delta e e^{-1})\updown{i}{k} + N\downup{i}{k}(\delta e e^{-1})\updown{i}{l} (\delta e e^{-1})\updown{l}{k}),
\end{Align}
where we have used the definition of the charge matrix \eqref{eq:chargesNij}, and the identity $\mathrm{det}g \, g^{jk}= \epsilon^{jm} \epsilon^{kn}g_{mn}$. This expression can be
put in a more familiar form. Introducing the pairing between matrices $\langle A,B \rangle = A\downup{i}{j}B\updown{i}{j}$, we can write $\Omega_\mathcal{R}^X = \delta \Theta^X$ with
\begin{equation}\label{eq:liouvillesymplecticpotential}
    \Theta^X = \sigma_S^{\mathcal{R}}\langle N_r, e^{-1}\delta e \rangle,
\end{equation}
which is the Liouville ``$p\,\dd q$'' one-form of $\rm T^*G$, with $e\in G$.
Since $N_r$ is traceless, the potential does not depend on the determinant of $e$
\begin{equation}
    \mathrm{tr}(e^{-1}\delta e)=\delta\ln\det e.
\end{equation}

The corner theory of the edge mode is therefore expressed by the standard symplectic structure of
$\rm T^*\spl{2}$ where the canonical variables are given by the group valued frame $e$ and the momentum is given by the relational Noether charge $N_r$.
In that sense, the edge mode can be thought of as a dynamical reference frame. Before moving to the quantum version, let us look at how symmetries act on the frame. For a reference space diffeomorphism $X \mapsto X\circ z$, the frame transforms as
\begin{equation}
    e\updown{i}{a} \xmapsto{z} e\updown{i}{b}\,\partial_a z^b.
\end{equation}
Those are the frame reorientations. They act on the reference space index on the right of the frame.
As they do not move the fields the charge is invariant whereas the relational charge transforms as
\begin{equation}
    N_r \xmapsto{z} h^{-1}N_r h, \qquad h = \partial z\eval_{S_r}.
\end{equation}
For a spacetime diffeomorphism $\psi$ fixing the corner, acting as $X \mapsto \psi^{-1} \circ X$ and $\varphi\mapsto\psi^*\varphi$, the frame transforms as
\begin{equation}\label{eq:frametransfo}
    e\updown{i}{a}\xmapsto{\psi} (\Lambda^{-1})\updown{i}{j}\, e\updown{j}{a},\qquad \Lambda=\partial\psi\eval_{S}.
\end{equation}
Spacetime diffeomorphisms thus act on the spacetime index on the left. Since the fields also transform $\varphi \mapsto \psi^* \varphi$, the charge now transforms while the relational charge is held fixed 
\begin{equation}\label{eq:chargetransfo}
    N \xmapsto{\psi} \Lambda^{-1} N \Lambda.
\end{equation}
A priori, $h,\Lambda$ are elements of the general linear group. However, it is easy to see that their determinant acts trivially on the frame. This is consistent with the fact that the determinant of the frame is gauge and does not appear in the symplectic form. The frame reorientations $h$ and spacetime diffeomorphisms $\Lambda$ can therefore consistently be taken as elements of $\spl{2}$.
The transformations \eqref{eq:frametransfo} and \eqref{eq:chargetransfo} are the frame part and the field part of a single spacetime diffeomorphism. 
To see how their charges combine, we treat the momentum of the frame as an independent variable $p_{\mathcal R}$ 
and consider the standard Liouville potential on the cotangent bundle of $\spl{2}$
\begin{equation}
    \Theta^{\mathrm{T}^{*}\spl{2}}_{\mathcal{R}} = \sigma^{\mathcal R}_S\langle p_{\mathcal R},e_{\mathcal R}^{-1}\delta e_{\mathcal R}\rangle,
\end{equation}
where the brackets denote the pairing between the Lie algebra of $\spl{2}$ and its dual.
From this expression it is easy to see that $p_{\mathcal R}$ generates the right translations of the frame, and $e_{\mathcal R}p_{\mathcal R}e_{\mathcal R}^{-1}$ its left translations. 
For $\Lambda=\mathds{1}+\lambda$, with $\lambda=\partial\xi\eval_S$, the total diffeomorphism charge is obtained by adding the field charge in equation \eqref{eq:Qxi} to the frame charge arising from the above potential. For corner preserving diffeomorphisms, the resulting charge reads
\begin{equation}
    H_\xi^{\mathcal{R}} = \sigma_S^{\mathcal{R}} \langle \mathcal{G}_{\mathcal{R}},\lambda\rangle,
\end{equation}
with
\begin{equation}\label{eq:gausslaw}
 \mathcal G_{\mathcal R}\defeq N[\varphi_{\mathcal R}]\eval_S-e_{\mathcal R}\,p_{\mathcal R}\,e_{\mathcal R}^{-1},
\end{equation}
where $N[\varphi_{\mathcal R}]$ denotes the functional \eqref{eq:chargesNij} of the fields of region $\mathcal R$. Diffeomorphism invariance requires $\mathcal G_{\mathcal R}\approx0$. 
This is a gravitational Gauss law at the corner, the analogue of the boundary component of the Gauss law in Yang--Mills theory~\cite{Donnelly:2016auv}. 
Solving it gives $p_{\mathcal R}=N_r$, and the potential reduces to \eqref{eq:liouvillesymplecticpotential}. 
The relational charge $N_r=e^{-1}N[\varphi]\,e$ is therefore gauge invariant, while the left momentum of the frame is identified with the charge of the fields. 
This is precisely the construction of relational observables recalled in the introduction. 
Neither the corner geometry $N[\varphi]$ nor the frame $e$ is gauge invariant on its own, since both transform under spacetime diffeomorphisms, 
but the components of the corner geometry measured in the frame's basis are.
\par 
We now move to the quantum theory. We can quantize the cotangent bundle $\rm T^*G$ in the Schrödinger polarization, as for a particle on a group or the link variables of lattice gauge theory \cite{Kogut:1974ag,Donnelly:2011hn}.
This means that states are described by wavefunctions of the frame in $L^2(\spl{2})$, on which $p=N_r$ and $e\,p\,e^{-1}=N$ act as the generators of right and left translations respectively. For a single region, the Gauss law \eqref{eq:gausslaw} would contract the left factor of $L^2(\spl{2})$ with the corner representation of the fields, as in~\cite{Donnelly:2016auv}. After gluing, the field charge is instead eliminated by the continuity condition and the fields are fixed by a gauge choice, as discussed below. We therefore only represent the frames, with kinematical Hilbert space $L^2(\spl{2})$ for each region. More precisely,
we restrict the quantum space to the principal series that corresponds to real values of the area \eqref{eq:casimirvalueorbits}. As we will see in the next section, causality further restricts the
allowed quantum configuration, which will be crucial for the derivation of the entropy.
\par    
The principal series representations of $\spl{2}$ are given by
\begin{equation}
    \mathcal{P}_{k,\epsilon}=\overline{\mathrm{span}}\qty{\ket{k,\epsilon;m}\ \big|\ m\in\epsilon+\mathbb{Z}},
    \qquad
    \braket{k,\epsilon;m}{k,\epsilon;m'}=\delta_{mm'},
\end{equation}
labelled by $k\in[0,\infty)$ and the parity $\epsilon\in\qty{0,\frac12}$, on which the generators act as
\begin{equation}
    L_0\ket{k,\epsilon;m}=m\ket{k,\epsilon;m},
    \qquad
    L_\pm\ket{k,\epsilon;m}=\qty(m\pm\tfrac12\pm ik)\ket{k,\epsilon;m\pm1}.
\end{equation}
These operators satisfy $[L_0,L_\pm]=\pm L_\pm$, $[L_-,L_+]=2L_0$, $L_0^\dagger=L_0$ and $L_\pm^\dagger=L_\mp$, and the quadratic Casimir takes the value
\begin{equation}
    \mathcal{C}=L_0^2-\frac12\qty(L_+L_-+L_-L_+)=-\qty(\frac14+k^2).
\end{equation}
Under the moment map provided by the Noether charges \eqref{eq:noethercharges}, the generator $w_1$ is represented by $D = \frac{i}{2}(L_+- L_-)$. If we interpret the
principal series as the quantization of the hyperbolic coadjoint orbit described by \eqref{eq:casimirvalueorbits}, the identification $k=\Phi_h$ still holds. Note that if one
were to have a different prescription for the identification of the representation parameter, it would only impact the subleading terms of the entropy computed in the next section.
\par   
Let us now consider our two subregions again. In order to ease into the construction, we start by describing the factorization for states that live in a given sector of the representations. This will be generalized later. For each subregion $\mathcal{R}\in\qty{\rm I,\rm{II}}$, the corner is described by a copy of the principal series, $\mathcal{H}_{\mathcal{R}}\equiv\mathcal{P}_{k_{\mathcal{R}},\epsilon_{\mathcal{R}}}$.
Naively, one could take the Hilbert space describing the entangling surface as 
the tensor product of those two spaces which would correspond to the statement that the two systems are independent. 
As discussed in the introduction, the gauge constraints relate the data between the two Cauchy slices and the naïve factorization fails. The actual
Hilbert space describing the entangling surface should instead be a proper subspace of the tensor product
\begin{equation}
    \mathcal{H}_G \subset \mathcal{H}_{\rm I} \otimes \mathcal{H}_{\rm II}.
\end{equation}
In \cite{Donnelly:2016auv}, it was proposed that this glued Hilbert space should be constructed via an entangling product that imposes the gauge constraint at the corner. 
Once the two subregions are glued, $S$ is no longer a boundary, and the charges of a simultaneous relabelling of the two reference corners must add up to zero, $H^{\rm I}_w+H^{\rm II}_w=0$. As in~\cite{Donnelly:2016auv}, this diagonal relabelling is therefore gauge. 
Denoting by $U_{\mathcal R}$ the unitary representation on $\mathcal H_{\mathcal R}$ generated by the charges $H^{\mathcal R}_w$, this is represented on the tensor product by the diagonal action
   \begin{equation}
         \mathcal{U}(h)=U_{\mathrm I}(h)\otimes U_{\mathrm{II}}(h),\qquad h\in\spl{2} .
   \end{equation}
   The entangling product is the subspace of invariant states,
   \begin{equation}
       \mathcal{H}_G=\mathcal{H}_{\rm I}\otimes_{\spl{2}}\mathcal{H}_{\rm II}=\qty{\ket{\psi}\in\mathcal{H}_I\otimes\mathcal{H}_{II}\ \big|\  \mathcal{U}(h)\ket{\psi}=\ket{\psi}\ \forall h\in\spl{2}},
   \end{equation}
   which is the quantum version of the gluing constraint $H^I_w+H^{II}_w=0$. By Schur's lemma, such invariant states exist only if $U_{\rm{II}}$
is equivalent to the contragredient of $U_{\rm I}$, and they are then unique up to normalization.\footnote{Since $\mathcal{P}_{k,\epsilon}$ is infinite-dimensional, the identity is not a Hilbert--Schmidt operator, so that $\mathcal{H}_{\rm I}\otimes\mathcal{H}_{\rm II}$ contains no normalizable invariant vector. As usual for non-compact gauge groups, \eqref{eq:maximallyentangledstate} is understood as a generalized vector, i.e.\ an element of the dual of a dense subspace in the rigged Hilbert space sense. Its divergent norm is the origin of the regularization discussed in section~\ref{sec:entanglemententropy}.} The invariant is therefore the identity intertwiner
written as a vector
\begin{equation}\label{eq:maximallyentangledstate}
    \ket{\mathds{1}}_{k,\epsilon}=\sum_{m\in\epsilon+\mathbb{Z}}\ket{k,\epsilon;m}_{\rm I}\otimes\ket{\overline{k,\epsilon;m}}_{\rm II},
\end{equation}
where $\ket{\overline{k,\epsilon;m}}_{\rm II}$ denotes the image of $\ket{k,\epsilon;m}$ under the identification of $\mathcal{H}_{\rm II}$ with the conjugate of $\mathcal{H}_I$.
Since the principal series is self-contragredient, the gluing condition
reduces to $k_{\rm I}=k_{\rm II}$ and $\epsilon_{\rm I} = \epsilon_{\rm II}$.
\par   
Now that we understand the entangling product in one sector of the principal series, we are ready to treat the full quantum reference frame.
As we go forward, we will compare the construction to the equivalent formulation for compact group, which might be more familiar to the reader. For $G$ compact, the Peter--Weyl theorem gives the harmonic decomposition of the $L^2$ space
\begin{equation}
    L^2(G)=\bigoplus_R\mathcal{H}_R\otimes\mathcal{H}_R^* ,
    \qquad
    \delta(g)=\frac{1}{\mathrm{Vol}(G)}\sum_R d_R\,\chi_R(g),
\end{equation}
where the second equation is the decomposition of the Diract distribution (defined with respect to the Haar measure) at the identity element of the group.
In the continuous case arising from non-compact groups, the generalization of this result is the Plancherel theorem. Restricting to the principal series
(the full decomposition of $L^2(\spl2)$ also contains the discrete series), it reads
\begin{equation}\label{eq:planchereltheorem}
    L^2(\spl{2})\supset\sum_{\epsilon}\int^\oplus\dd\mu_\epsilon(k)\,\mathcal{P}_{k,\epsilon}\otimes\mathcal{P}_{k,\epsilon}^* ,
    \qquad
    \delta(g)\supset\sum_{\epsilon}\int\dd\mu_\epsilon(k)\,\chi_{k,\epsilon}(g),
\end{equation}
where the Plancherel measure $\dd\mu_\epsilon(k)=\mu_\epsilon(k)\,\dd k$ is fixed by the group structure (the invariance of the Haar measure) up to an overall constant. The second relation is the Plancherel inversion formula, which expands the Dirac
distribution at the identity element in characters
$\chi_{k,\epsilon}$.
It shows that the Plancherel density weighs each sector as the dimension $d_R$ does in the compact case.
The quantum reference frames associated with the edge modes are thus described by wavefunction 
$\Psi(g)\in\mathbb{C}$, $g\in \spl{2}$, i.e.\ elements of $L^2(\spl{2})$, on which 
the surface symmetries act by right translations, $\psi(g)\to\psi(gh)$.
Using the Plancherel decomposition, any wavefunction can be written
\begin{equation}
        \Psi(g) = \sum_\epsilon\int \dd\mu_\epsilon(k)\Tr\qty(U_{k,\epsilon}(g)\hat\Psi_\epsilon(k)), \qquad \hat\Psi_\epsilon(k) = \int \dd g\, \Psi(g)\,U_{k,\epsilon}(g^{-1}),
\end{equation}
where $\dd g$ is the Haar measure on the group.
\par 
On the Fourier coefficients, frame transformations act as $\hat\Psi_\epsilon(k)\to U_{k,\epsilon}(h)\hat\Psi_\epsilon(k)$. These therefore act
on the $\mathcal{P}_{k,\epsilon}$ space (on the ket side). Conversely, the frame part \eqref{eq:frametransfo} of spacetime diffeomorphisms acts via left translations $\Psi(g)\to\Psi(\Lambda^{-1}g)$, which corresponds to
$\hat\Psi_\epsilon(k)\to\hat\Psi_\epsilon(k)\,U_{k,\epsilon}(\Lambda)^{-1}$. These therefore act
on the $\mathcal{P}^*_{k,\epsilon}$ space (on the bra side).
The glued Hilbert space of \cite{Donnelly:2016auv} is obtained by requiring the two-sided wavefunctions $\Psi(g_{\rm I},g_{\rm II})$ to be invariant under the frame transformation's diagonal action
\begin{equation}\label{eq:rightinvariance}
    \Psi(g_{\rm I}h,g_{\rm II}h)=\Psi(g_{\rm I},g_{\rm II})\qquad\forall h\in \spl{2},
\end{equation}
so that glued wavefunctions depend only on the relative frame, $\Psi(g_{\rm I},g_{\rm II})=F(g_{\rm I}g_{\rm II}^{-1})$. For $G=\mathbb{R}$, this is
the statement that translation-invariant two-particle states depend only on the relative position of the particles. In our case, this is the statement
that the reference labels are conventions. In other words, relabelling both frames at once changes nothing physical. In terms of charge constraints, this corresponds to
$N_r^{\rm I} = N_r^{\rm II}$, which means that both frames report the same numbers for the corner geometry.
\par    
Let us now study the gauge transformations. For a single region, every corner-preserving diffeomorphism is gauge, since its charge \eqref{eq:gausslaw} vanishes. However, gluing the two regions adds a condition. The glued spacetime carries a single smooth metric and dilaton, so that the field charges of the two regions, which are both tensors on the same tangent plane $\rm{T}_SM$, must agree
\begin{equation}\label{eq:continuity}
    \mathcal C\defeq N[\varphi_{\rm I}]\eval_S-N[\varphi_{\rm II}]\eval_S\approx0 .
\end{equation}
A diffeomorphism of region $\rm I$ alone, whose Jacobian at $S$ does not commute with $N$, rotates the 
corner geometry of one side only and violates \eqref{eq:continuity}. In the language of constrained systems, 
the one-sided Gauss constraints become second class, except for their component along $N$, which generates 
a boost of the normal plane and preserves the corner geometry (Appendix~\ref{app:constraints}). Solving 
the second-class constraints identifies the left momenta of both frames with the field charge, 
$e_{\rm I}p_{\rm I}e_{\rm I}^{-1}=e_{\rm II}p_{\rm II}e_{\rm II}^{-1}=N[\varphi]\eval_S$, without 
changing the Poisson brackets of the frames. The remaining diagonal Gauss constraint acts on the 
fields through the sum of their charges, which cancels since $\sigma^{\rm I}_S=-\sigma^{\rm II}_S$, 
and therefore only acts on the frames, as their diagonal left translation. The Yang--Mills case 
of~\cite{Donnelly:2016auv} is different.\footnote{In~\cite{Donnelly:2016auv}, Yang--Mills gauge transformations act on the trivialisation $\phi$ from the right and surface symmetries from the left. In terms of $e=\phi^{-1}$, the left action denotes gauge transformations in both theories.} 
There, the gluing condition only involves the dressed charges, which are invariant under the gauge 
transformations of either region. It commutes with both one-sided Gauss constraints, which 
remain first class, so that gauge transformations remain gauge independently on each side. 
The difference has a geometric origin. In Yang--Mills theory, the bulk fields of each region 
are expressed in a local trivialisation of that region, and a one-sided gauge transformation is a 
change of this trivialisation. In gravity, the bulk fields are tensors on the single manifold $M$, 
and the corresponding operation is a one-sided diffeomorphism, which does not preserve the smoothness 
of the fields at $S$ unless it acts there as a boost. At the quantum level, the diagonal Gauss constraint 
requires
\begin{equation}\label{eq:leftinvariance}
   \Psi(\Lambda g_{\mathrm I},\Lambda g_{\mathrm{II}})=\Psi(g_{\mathrm I},g_{\mathrm{II}})\qquad\forall \Lambda\in \spl{2}.
\end{equation}
Classically, this is the constraint $N^{\rm I}=N^{\rm II}$ on the left momenta $N^{\mathcal R}\defeq e_{\mathcal R}N_r^{\mathcal R}e_{\mathcal R}^{-1}$ of the two frames, which both equal the field charge at $S$. Since the one-sided left actions are no longer gauge, except for the boost discussed below, the relative frame $\mathsf{r}=e_{\rm {I}}\,e_{\rm {II}}^{-1}\in\spl{2}$, 
which maps the axes of frame $\rm{II}$ onto those of frame $\rm{I}$, is no longer 
pure gauge. In order to see which degree of freedom survives, we write
\begin{equation}
    N^{\rm I} = e_{\rm I} N_r e^{-1}_{\rm I} =\mathsf{r}\, e_{\rm {II}}N_re^{-1}_{\rm {II}}\,\mathsf{r}^{-1} = \mathsf{r}\,N^{\rm{II}}\mathsf{r}^{-1},
\end{equation}
where we have used the first condition, $N_r = N_r^{\rm I} = N_r^{\rm{II}}$. The 
second condition, $N = N^{\rm I} = N^{\rm{II}}$, then gives
\begin{equation}\label{eq:rNr}
    \mathsf{r}\,N\, \mathsf{r}^{-1} = N .
\end{equation}
The relative frame must therefore commute with the charge. Using the value \eqref{eq:Ncharges} of the charge at the bifurcation point in Kruskal--Szekeres coordinates, which can always be reached with the diagonal action, we find
\begin{equation}
    \mathsf{r} = \mqty(\cosh\eta&\sinh\eta\\ \sinh\eta&\cosh\eta),
\end{equation}
up to an overall sign that reverses the time orientation and is excluded 
by causality. This is a boost of rapidity $\eta$ of the normal plane. The two frames can therefore only differ by a boost along 
the Killing flow as any other relative transformation would move the null 
directions, and the two sides would disagree on the location of the 
horizons. Equivalently, since $N$ is built from $\hg$ and $\Phi$ only, its stabilizer in $\spl{2}$ is the group of isometries of $\hg$ at $S$, and \eqref{eq:rNr} states that the two frames induce the same metric on the reference normal plane, $e_{\rm I}^{T}\,\hg\,e_{\rm I}=e_{\rm II}^{T}\,\hg\,e_{\rm II}$. The physical metric is smooth at $S$ for any $\mathsf r$ and the continuity condition is the statement that, pulled back to reference space, it remains continuous across $S_r$.
\par
The boost of the normal plane is precisely the exception noted above. The one-sided Gauss constraint along 
$N$ remains first class, and the corresponding diffeomorphism of region $\rm I$ is gauge. It can be 
chosen to coincide with the Killing boost near $S$, where it leaves the fields invariant and 
shifts $\eta$. Since it vanishes far from $S$, it also deforms the fields of region $\rm I$ where 
it is cut off. This deformation shifts the relative Killing time $\tau$ of the two asymptotic ends 
along $\Sigma$, the variable conjugate to the mass found by Kucha\v{r}~\cite{Kuchar:1994zk}, so that only 
the combination $\eta-\kappa\tau$ is gauge invariant, up to orientation conventions 
(Appendix~\ref{app:constraints}). We use this freedom to set $\tau=0$. Together with the diffeomorphisms that are smooth at $S$, which bring the fields to the Kruskal--Szekeres form of Section~\ref{sec:dimensionallyreducedblackhole}, this selects the configuration in which $\Sigma$ is the straight slice of Figure~\ref{fig:blackhole}.
The fields then form a one-parameter family labelled 
by the mass, on which $\Omega_{\rm I}+\Omega_{\rm II}$ vanishes, and the frames carry the whole symplectic 
structure. This is not the only possible gauge. Using instead the one-sided boost to set $\eta=0$, and the 
diagonal action to set $e_{\rm I}=e_{\rm II}=\mathds{1}$, fixes the coordinate systems completely. The fields can still be brought to Kruskal--Szekeres form, but $\Sigma$ is then bent with respect to the surfaces of constant Killing time, by an amount measured by $\tau$. 
This is the unitary gauge of~\cite{Speranza:2022lxr}. Since $\delta e_{\mathcal R}=0$, the corner terms of \eqref{eq:odfin2d} drop out, 
the extended symplectic form reduces to $\Omega_{\rm I}+\Omega_{\rm II}$, and the pair 
is $(A/8\pi G_N,\tau)$, as in the analysis of Kucha\v{r}, which involves no frames. 
As shown in~\cite{Speranza:2022lxr}, such a gauge exists because $\Omega^{\rm DF}_{\mathcal R}$ is 
degenerate along all diffeomorphisms. The coordinate systems then act as Stueckelberg fields, and 
the extended phase space of~\cite{Donnelly:2016auv} adds no physical degree of freedom to that of the 
fields. The two descriptions are thus the same reduced phase space in two gauges, $\tau=0$ and $\eta=0$, 
the gauge-invariant variable $\eta-\kappa\tau$ being carried by the frames in the first and by the fields 
in the second. They differ however in how the Hilbert space factorizes across $S$, on which the 
entanglement entropy depends. In the gauge $\tau=0$, the pair is carried by the relative orientation of 
two frames, each attached to one region and acted on by $\spl{2}$, and this is the gauge in which we 
quantize.
\par
In this gauge, the single physical degree of freedom carried by the glued corner
(besides the area) is thus the relative rapidity $\eta$ between the two
frames, i.e.\ the hyperbolic opening angle between the two sides at the 
bifurcation point. This is precisely the variable identified by Carlip and 
Teitelboim \cite{Carlip:1993sa}, who showed that the opening angle 
$\Theta$ at the bifurcation surface is canonically conjugate to 
$A/8\pi G_N$.
This can be seen even more explicitly at the classical level. There, the two gluing conditions are the moment map constraints of the diagonal right and left actions, and the gluing is the corresponding symplectic reduction of the two copies of $T^*G$. Imposing $N_r^{\rm I}=N_r^{\rm II}$ and writing $e_I=\mathsf{r}\,e_{II}$, the sum of the potentials \eqref{eq:liouvillesymplecticpotential}
becomes
\begin{equation}
    \Theta^X_{\rm I}+\Theta^X_{\rm II}=\sigma^{\rm I}_S\,\langle N^{\rm II},\mathsf{r}^{-1}\delta\mathsf{r}\rangle ,
\end{equation}
where we have used $\sigma_S^{\mathrm{I}} = - \sigma_S^{\mathrm{II}}$. After gluing from the right, we are therefore left with only one copy of $T^*G$, for the relative frame with momentum $N$. The left gluing then imposes \eqref{eq:rNr} and quotients the conjugation $\mathsf{r}\to\Lambda^{-1}\mathsf{r}\,\Lambda$.
Using this freedom to bring the charge to its value \eqref{eq:Ncharges}, $N=\Phi_h J$ with $J_T{}^R=J_R{}^T=1$, the relative frame is the boost $\mathsf{r}=e^{\eta J}$ and
\begin{equation}\label{eq:gluedOmega}
    \Theta^X_{\rm I}+\Theta^X_{\rm II}=2\sigma^{\rm I}_S\,\Phi_h\,\delta\eta ,
    \qquad
    \Omega=2\sigma^{\rm I}_S\,\delta\Phi_h\doublewedge\delta\eta,
\end{equation}
where the factor of two arises from $\langle J,J\rangle=2$ and where we briefly reinstated the exterior product between field space forms, denoted by $\doublewedge$. The glued corner phase space is thus two-dimensional: the area $2\Phi_h=A/8\pi G_N$ and the relative rapidity $\eta$ form a single canonical pair, which is the one of Carlip and Teitelboim.
That a single canonical pair survives is not a coincidence. By Birkhoff's theorem, 
every spherically symmetric vacuum solution is locally static, so that the bulk 
carries no local degrees of freedom and the reduced phase space of the two-sided system is 
two-dimensional~\cite{Kuchar:1994zk}. The fields alone already carry this pair, $(A/8\pi G_N,\tau)$, 
since a boost acting on the fields of one region only has the non-vanishing charge 
$\sigma^{\rm I}_S\langle N,\lambda\rangle$. The frames do not add a second pair, since the one-sided boost 
identifies $\eta$ with $\kappa\tau$. Gluing the frames from the right only would instead leave the 
full relative frame $\mathsf r\in\spl{2}$ and its momentum, a six-dimensional phase space. 
The role of the continuity condition \eqref{eq:continuity} is different. Without it, the one-sided 
Gauss constraints would remain first class, as in Yang--Mills theory, the relative frame would be pure 
gauge, and the pair would be carried by the fields alone. With it, the pair is carried by the relative 
orientation of the two frames, on which $\spl{2}$ acts. This will be crucial for the entropy.
\par 
As anticipated in the introduction, this pair admits two further readings. The first relates it to the analysis of 
Kucha\v{r}~\cite{Kuchar:1994zk}. Since the solution is static, the one-sided boost that shifts $\eta$ is generated by the Killing vector. 
It extends to an isometry of the whole exterior region~$\rm I$. A one-sided boost of rapidity $\eta$ therefore
shifts the Killing time of region~$\rm I$ by $\eta/\kappa$. With the Killing vector 
normalized at infinity, this is the shift in the reading of a static clock at the asymptotic end of region~$\rm I$, which is the clock used by Kucha\v{r} in the asymptotically flat case.
In our phase space this shift is generated by the area, while in Kucha\v{r}'s 
reduced phase space, the shift of the clock at one infinity is generated by the mass 
$M$. The two generators are related by the first law, which equates the variations of the Noether charge of the Killing vector at the bifurcation point and at infinity~\cite{Iyer:1994ys}.
Accordingly, the change of variables $(A/8\pi G_N,\eta)\mapsto(M,T)$, with $M$ fixed by the first law and $T=\eta/\kappa$, is a canonical transformation onto Kucha\v{r}'s pair.\footnote{With $a\defeq A/8\pi G_N$ and $\kappa=\kappa(a)$ on the family of solutions, $\delta a\doublewedge\delta\kappa=0$, so that $\delta M\doublewedge\delta T=\kappa\,\delta a\doublewedge\big(\kappa^{-1}\delta\eta-\eta\,\kappa^{-2}\delta\kappa\big)=\delta a\doublewedge\delta\eta$. This was already observed in~\cite{Gour:2002pj}} 
The two edge frames thus play the role of Kucha\v{r}'s two asymptotic clocks, 
and the relative rapidity is his relative reading, transported to the corner along the Killing flow. 
Louko and Whiting~\cite{Louko:1994tv} interpolate between the two pictures: replacing one asymptotic end by the bifurcation sphere, 
the variable conjugate to the mass includes the boost parameter elapsed there, which they fix as boundary data, while here it is the dynamical relative orientation of the 
two frames. 
In the extended phase space, this identification is the equivalence between the gauges $\tau=0$ and $\eta=0$ discussed above. The frame-only boost, generated by the area, and the shift of the asymptotic Killing time, generated by the mass, differ by the one-sided boost cut off away from $S$, which is gauge. The asymptotic regions only enter through the normalization of the Killing vector, and we do not include Kucha\v{r}'s boundary terms. 
The second reading is that the relative rapidity is the timeshift of the two-sided crossed product construction~\cite{Chandrasekaran:2022eqq}, which is conjugate to the boundary energies and, through the first law, to the area. That the area generates a relative boost between the two sides of an extremal surface is the content of the kink transform~\cite{Bousso:2020yxi}. Here, this degree of freedom arises from the gluing of the two edge frames.
\par 
Getting back to the quantum case, the two gluing conditions can be expressed as the condition that the glued Hilbert space wavefunctions are given by class functions of the relative frame
\begin{equation}\label{eq:gluedstateisclassfunction}
   \Psi(g_{\mathrm I},g_{\mathrm{II}})=F(g_{\mathrm I}g_{\mathrm{II}}^{-1}), \qquad F(\Lambda g\Lambda^{-1})=F(g)\quad \forall \Lambda \in \spl{2}.
\end{equation}
By the Plancherel theorem, restricted to the principal series, class functions are superpositions of characters
\begin{equation}
    F(g)=\sum_\epsilon\int\dd\mu_\epsilon(k)\,\psi_\epsilon(k)\,\chi_{k,\epsilon}(g),\qquad \chi_{k,\epsilon}(g)=\Tr U_{k,\epsilon}(g).
\end{equation}
To interpret this expansion, recall that by the Plancherel theorem 
\eqref{eq:planchereltheorem}, the matrix elements 
$U_{ab}(g)=\bra{a}U_{k,\epsilon}(g)\ket{b}$ (where $a,b$ label an orthonormal basis of $\mathcal{P}_{k,\epsilon}$) form a basis of 
$\mathcal{P}_{k,\epsilon}\otimes\mathcal{P}^*_{k,\epsilon}$. 
Writing the group law at the level of the matrix elements
\begin{equation}
    U_{ab}(gh)=\sum_c U_{ac}(g)U_{cb}(h),\qquad U_{ab}(\Lambda^{-1}g)=\sum_c U_{ac}(\Lambda^{-1})U_{cb}(g),
\end{equation}
we see that 
frame transformations act on the index $b$ and spacetime diffeomorphisms 
on the index $a$. Using unitarity, the character of the relative frame 
reads
\begin{equation}
    \chi_{k,\epsilon}(g_{\rm I}g_{\rm II}^{-1})=\Tr\qty[U_{k,\epsilon}(g_{\rm I})U_{k,\epsilon}(g_{\rm II})^{-1}]=\sum_{a,b}U_{ab}(g_{\rm I})\,\overline{U_{ab}(g_{\rm II})}.
\end{equation}
As a two-sided wavefunction, it contracts the frame index of $\mathrm{I}$ with the frame index of $\mathrm{II}$, and the spacetime index of $\mathrm{I}$ with the spacetime index of $\mathrm{II}$. Each contraction is an identity intertwiner of the form \eqref{eq:maximallyentangledstate}, so that the glued states read
\begin{equation}\label{eq:doublygluedstate}
    \ket{\psi}=\sum_\epsilon\int\dd\mu_\epsilon(k)\,\psi_\epsilon(k)\;\ket{\mathds{1}_{k,\epsilon}}_{\rm frame}\otimes\ket{\mathds{1}_{k,\epsilon}}_{\rm spacetime}.
\end{equation}
The first intertwiner is the state obtained from the gluing of \cite{Donnelly:2016auv} which expresses the fact 
that both sides agree on the reference frame. The second is its analogue for the spacetime factors, enforced by the continuity 
condition. It expresses the fact that both sides agree on the corner geometry, without either of them carrying an absolute orientation 
of the corner in spacetime. Equation~\eqref{eq:doublygluedstate} is the gravitational analogue of the compact-group state~\eqref{eq:superposition} discussed below, with a second intertwiner that has no counterpart in Yang--Mills theory, where the spacetime factor of each frame is instead contracted with the bulk fields of its own region.
\section{Entanglement Entropy}\label{sec:entanglemententropy}
Now that the entangling product has been constructed, we can discuss the associated entanglement entropy. The main difficulty lies in the
infinite dimensionality of the representations of the non-compact group. At the state level, this translates into the fact that \eqref{eq:doublygluedstate} is not normalizable.
Consequently, the density operator $\rho = \ket\psi \bra \psi$ is not trace class and its von Neumann entropy diverges. 
We will show that this divergent term only contributes to a state-independent constant. Note that this is a familiar situation in the study of type II von Neumann algebras.
We start this section with a reminder of the derivation of the entanglement entropy associated with the entangling product in the case of a compact group.
We then move to the generalization to the non-compact case of interest. Next, we introduce the concept of causal frames, which restricts the allowed frames to the semigroup $\mathrm{SL}^+(2,\mathbb{R})$. We finally use the Plancherel measure of the semigroup to compute
the entanglement entropy and show that the dominant contribution in the classical limit is the Bekenstein--Hawking area term.
\par    
We first recall how the entangling product determines the entanglement entropy \cite{Donnelly:2016auv}. It is instructive to start with a group whose irreducible representations $R$ are finite-dimensional, with dimension $d_R$. The corner Hilbert space of each subregion is a direct sum over superselection sectors, $\mathcal{H}_{\mathcal{R}}=\bigoplus_R\mathcal{H}_{\mathcal{R},R}$, and the gluing acts sector by sector. In a given sector, the glued space is spanned by the single invariant state
\begin{equation}
    \ket{\mathds{1}_R}=\frac{1}{\sqrt{d_R}}\sum_{m=1}^{d_R}\ket{R;m}_{\mathrm{I}}\otimes\ket{\overline{R;m}}_{\mathrm{II}},
\end{equation}
which is now normalized. A general glued state is a superposition of sectors
\begin{equation}\label{eq:superposition}
    \ket{\psi}=\sum_R\psi(R)\ket{\mathds{1}_R},\qquad \sum_R\abs{\psi(R)}^2=1,
\end{equation}
and tracing over $\mathcal{H}_{\rm II}$ gives the reduced density matrix
\begin{equation}\label{eq:rhoI}
    \rho_{\mathrm{I}}=\Tr_{\mathrm{II}}\ket{\psi}\bra{\psi}=\bigoplus_R\,p(R)\,\frac{\mathds{1}_{R}}{d_R},\qquad p(R)=\abs{\psi(R)}^2.
\end{equation}
Where $\mathds{1}_R$ is the unit operator in the Hilbert space $\mathcal{H}_{\mathrm{I},R}$. The off-diagonal terms between different sectors drop out because the two sides are correlated sector by sector. Hence $\rho_{\mathrm{I}}$ is block diagonal, and within each block it is maximally mixed. Its von Neumann entropy is
\begin{equation}\label{eq:DFentropy}
    S=-\Tr\rho_{\mathrm{I}}\ln\rho_{\mathrm{I}}=-\sum_R p(R)\ln p(R)+\sum_R p(R)\ln d_R.
\end{equation}
The first term is the standard Shannon entropy of the distribution over sectors. The second can be understood as the entropy of the corner degrees of freedom themselves: in each sector the gauge constraint entangles the two copies of $\mathcal{H}_R$ maximally, contributing $\ln d_R$. Only the second term can scale with the size of the corner, since the first is bounded by the spread of $p(R)$.
We also note that the entropy can be computed from the standard limit of Rényi entropies 
\begin{equation}\label{eq:tracerhoncompact}
    S = \lim_{n\to 1}\frac{1}{1-n}\ln\Tr\rho_{\rm I}^n, \qquad     \Tr\rho_{\rm I}^n=\sum_R (d_R)^{1-n}p(R)^n.
\end{equation}
In what follows, we will find a direct analogue of the trace formula for $\rho_{\mathrm{I}}^n$ in the non compact case.
\par
Since the glued state is invariant under both diagonal actions, the reduced density matrix $\rho_I$ commutes 
with the right and left translations of the frame of region $\rm I$. By Schur's lemma, it acts in each sector as 
$\abs{\psi_\epsilon(k)}^2$ times the identity on $\mathcal{P}_{k,\epsilon}\otimes\mathcal{P}^*_{k,\epsilon}$. 
Here lies the first consequence of our gluing of the spacetime factor: the frame factor and the spacetime factor of region $\mathrm{I}$ are both maximally mixed.
With the frame gluing of \cite{Donnelly:2016auv} alone, as in Yang--Mills theory, the spacetime factor of each frame would instead be contracted with the corner data of the fields of its own region, and only the frame factor would be maximally mixed across $S$, as in the compact case \eqref{eq:rhoI}. 
On $L^2(\spl{2})$, the integral kernel of the reduced density operator is of the form $\rho_{\rm I}(g,g')=K(gg'^{-1})$, with $K$ a class function.\footnote{From equation \eqref{eq:gluedstateisclassfunction}, we can write
\begin{equation}
  \rho_{\mathrm{I}}(g,g') = \int_G \dd g_{\mathrm{II}}\, F(g g^{-1}_{\mathrm{II}})\overline{F(g' g^{-1}_{\mathrm{II}})}
     = \int_G \dd u\, F(g g'^{-1}u)\overline{F(u)} \equiv K(gg'^{-1}),
\end{equation}
where we used the invariance of the Haar measure under inversion and translations for unimodular $G$ such as $\spl{2}$. The left gluing then implies that $K$ is a class function
\begin{equation}
    K(\Lambda x \Lambda^{-1}) = \int_G \dd u\, F(\Lambda x u\Lambda^{-1})\overline{F(\Lambda u \Lambda^{-1})} = \int_G \dd u\, F(xu)\overline{F(u)} = K(x).
\end{equation}
The diagonal is thus constant, and the kernel of $\rho^n_{\mathrm{I}}$, which is the $n$-fold kernel $K^{(n)}(gg'^{-1})$ is still a class function. Equation \eqref{eq:trrhoIn} then follows.
}
Its diagonal is thus constant and we can write
\begin{equation}\label{eq:trrhoIn}
    \Tr\rho_{\rm I}^n=\mathrm{Vol}(G)\,K^{(n)}(e)=\mathrm{Vol}(G)\sum_\epsilon\int\dd\mu_\epsilon(k)\,\dim\mathcal{P}_{k,\epsilon}\,\abs{\psi_\epsilon(k)}^{2n},
\end{equation}
where $K^{(n)}$ is the kernel of $\rho_{\rm I}^n$, and the second equality follows from the Plancherel formula $f(e)=\sum_\epsilon\int\dd\mu_\epsilon(k)\Tr\hat f_\epsilon(k)$.
This expression contains two divergent factors. The first is that the volume of the group is infinite. As this is the same for every representation, it is an overall
divergent term that, as we will see later, only shifts the entropy by an infinite term. 
The second infinity lies in the dimension of each representation $\dim\mathcal{P}_{k,\epsilon}$. Here, we can extract
the divergent term by using the procedure described in Appendix B of \cite{Mertens:2022ujr}, where the authors argue that the dimension can formally be written
\begin{equation}\label{eq:Dregularization}
    \dim \mathcal{P}_{k,\epsilon} = 2\pi\frac{\mathrm{Vol}(G)}{V_C} \mu_{\epsilon}(k),
\end{equation}
where $V_C$ is the volume of a Cartan subgroup.
With this factorization of the divergent term, we can write the trace formula as 
\begin{Align}\label{eq:410}
    \Tr \rho_{\mathrm I}^n &=\frac{2\pi \mathrm{Vol}(G)^2}{V_C} \sum_\epsilon \int \dd k\, \mu_{\epsilon}^2(k)\abs{\psi_\epsilon(k)}^{2n}\\
    &= \sum_\epsilon \int \dd k\, \qty(\frac{2\pi \mathrm{Vol}(G)^2}{V_C}\mu^2_\epsilon(k))^{1-n} p_{\epsilon}(k)^n,
\end{Align}
where we defined
\begin{equation}
    p_\epsilon(k) \defeq \frac{2\pi}{V_C} \qty(\mathrm{Vol}(G) \mu_\epsilon(k) \abs{\psi_{\epsilon}(k)})^2.
\end{equation}
If $p_\epsilon(k)$ is normalized
\begin{equation}
    \sum_\epsilon \int \dd k\, p_\epsilon(k) = 1,
\end{equation}
this is the continuous analogue of equation \eqref{eq:tracerhoncompact}. We can then deduce the entanglement entropy of the edge mode from the Rényi formula
\begin{equation}\label{eq:finalentropy}
    S = -\sum_\epsilon \int \dd k\, p_\epsilon(k)\ln\frac{p_\epsilon(k)}{\mu_\epsilon(k)^2} + \ln\frac{2\pi \mathrm{Vol}^2(G)}{V_C}.
\end{equation}
Since $p_\epsilon$ is normalized, the divergence is entirely contained in the second term. As previously mentioned, the entanglement entropy being defined up to a divergent constant is familiar in type II von Neumann algebras and we discard it in what follows.
Each of the two gluing conditions contributes one factor $\ln\mu_\epsilon(k)$, one through 
the Plancherel measure and one through the regularized dimension \eqref{eq:Dregularization}. 
This will prove crucial in order to obtain the right numerical factor of 
the area law. The first term of equation~\eqref{eq:finalentropy} is the continuous analogue of \eqref{eq:DFentropy}.
\par
To compute the entanglement entropy of a state specified by $p_\epsilon(k)$, we therefore only need the Plancherel measure of the corner symmetry group.
As already mentioned, since we only consider real values of the area, we should restrict our attention to the principal series.
The Plancherel measure of the principal series is~\cite{HarishChandra:1952,Kitaev:2017hnr}
\begin{equation}
    \mu_0(k)\propto k\tanh(\pi k),\qquad \mu_{\frac12}(k)\propto k\coth(\pi k),
\end{equation}
Both densities grow only linearly at large $k$, such that the entanglement entropy in that regime grows like the logarithm of $\Phi_h$. As we will now see, this changes once causality is imposed, which restricts the allowed frames to a semigroup. Note that if one were to add the discrete series, they only contribute to the entanglement entropy
to logarithmic order as well.
\par
The edge frame \eqref{eq:edgeframe} maps the reference normal plane $\rm{T}_{S_r}U_r$ to the physical normal plane $\rm{T}_SM$, which carries the light cone of the metric. In the null coordinates $(U,V)$, the two-dimensional metric reads
\begin{equation}
    \hg_{ij}\dd x^i\dd x^j = - e^{2\hat\sigma}\, \dd U\, \dd V, \qquad e^{2\hat \sigma} = -\frac{\rho \ell^2}{(4G_N)^{\frac32}}\,\frac{f(\rho)}{\kappa^2 U V}.
\end{equation}
where the conformal factor is positive and smooth on the whole maximally extended spacetime. A vector $v\in T_SM$ is therefore causal if and only if $v^Uv^V\ge0$, independently of the conformal factor, and the closed future light cone of the corner is
\begin{equation}
    C_h=\qty{\alpha\,\partial_V+\beta\,\partial_U\ \big|\ \alpha,\beta\ge0}.
\end{equation}
In order to speak about causality, the reference plane must be equipped with a causal structure as well. We choose reference null coordinates $y^\pm$ and the reference future cone $C_0=\qty{v^+\partial_++v^-\partial_-\,|\,v^\pm\ge0}$. Like the reference coordinates themselves, this is part of the reference structure. We then require the frame to be causal
\begin{equation}\label{eq:causalframe}
    e(C_0)\subseteq C_h ,
\end{equation}
i.e.\ every future-directed causal reference direction is mapped to a future-directed causal physical direction. Physically, this corresponds to requiring that the reference frame is made of physical observers defined by a timelike reference direction with a notion of future compatible with that of the spacetime.
Since a cone is generated by its two edges, \eqref{eq:causalframe} holds if and only if the two reference null directions are mapped into $C_h$. Writing the frame in the two null bases
\begin{equation}
    e=\mqty(a&b\\c&d),\qquad e(\partial_+)=a\,\partial_V+c\,\partial_U,\qquad e(\partial_-)=b\,\partial_V+d\,\partial_U,
\end{equation}
the causality condition is simply $a,b,c,d\ge0$. Causal frames are therefore associated with the subset
\begin{equation}
    \mathrm{SL}^{\ge0}(2,\mathbb{R})=\qty{\mqty(a&b\\c&d)\in\spl{2}\ \Big|\ a,b,c,d\ge0}.
\end{equation}
Since the boundary of that set has a vanishing Haar measure, it can be dropped and we 
restrict our attention to the strictly positive entries whose set is denoted $\mathrm{SL}^+(2,\mathbb{R})$.
The product of two matrices with positive entries has positive entries, so that causal frames remain causal under the composition with causal transformations $e\to e\,h$ and $e\to \Lambda^{-1} e$ with $h,\Lambda^{-1}\in \mathrm{SL}^{+}(2,\mathbb{R})$.
Inverses, however, are generally not causal, and causal frames form a semigroup rather than a group. The relative frame $\mathsf r=e_{\rm I}e_{\rm II}^{-1}$ of two causal frames is in general not causal. After gluing, however, \eqref{eq:rNr} restricts it to a boost, $\mathsf r=\mathrm{diag}(e^{\eta},e^{-\eta})$ in the null basis, and boosts are precisely the elements of $\mathrm{SL}^{\ge0}(2,\mathbb{R})$ whose inverse is also causal.
Since causal frames preserve the future light cone, they act on the slopes $x$ of timelike lines through the origin by Möbius transformations 
that preserve the half-line $x>0$. The principal series of $\mathrm{SL}^+(2,\mathbb{R})$ is therefore 
realized on functions of $x>0$. As a consequence, it does not see the parity $\epsilon$,
leaving a single continuous series labelled by $k$.
\par   
The harmonic analysis on $\mathrm{SL}^+(2,\mathbb{R})$ is known from the positive representations of the modular double of $U_q(\mathfrak{sl}(2,\mathbb{R}))$ \cite{Ponsot:1999uf,Ponsot:2000mt} in the limit $q\to1$, as used in JT gravity \cite{Blommaert:2018oro,Blommaert:2018iqz}. The Plancherel measure is
\begin{equation}\label{eq:SLplusPlancherel}
    \dd\mu(k)\propto k\sinh(2\pi k)\,\dd k .
\end{equation}
Substituting the above in \eqref{eq:finalentropy} and removing the sum over $\epsilon$, we get
\begin{equation}
    S=-\int\dd k\,p(k)\ln p(k) + 2\int\dd k\,p(k)\ln(\mu(k))+\mathrm{const},
\end{equation}
where the constant term contains all divergences.
As mentioned earlier, the first term is the Shannon entropy of the distribution over the different sectors and the second is the edge-mode term.
Note that since the glued system carries no bulk degrees of freedom, both terms arise from the edge modes. Following the terminology of the gauge theory literature, we nevertheless reserve the name edge-mode term for the second one, which is the continuum analogue of the dimension term in \eqref{eq:DFentropy} and comes from the maximal entanglement of the two factors of the frame wavefunction within each sector. The first term only reflects the classical uncertainty in the area.
In what follows, we will therefore focus on this second term and call it $S_\mathrm{edge}$.
Note that a different overall factor in the normalization of the Plancherel measure simply shifts the entropy by a constant, we will therefore work with the equality sign in equation \eqref{eq:SLplusPlancherel}.
Let us consider a normalized profile $f(u)$, centered so that $\int u f(u) \dd u = 0$. We then consider the associated distribution centered on the classical value $k_* = \Phi_h$
\begin{equation}
    p(k) = \frac{1}{\Delta}f\qty(\frac{k-k_*}{\Delta}),
\end{equation}
where $\Delta$ is the width. For example, one could look at a Gaussian distribution $f(u) = e^{-\frac{u^2}{2}}\slash \sqrt{2\pi}$, but the result that follows covers the more general case.
The edge-mode entropy can then be written
\begin{Align}\label{eq:Sedgefull}
    S_{\mathrm{edge}} &= \frac{2}{\Delta} \int \dd k\,f\qty(\frac{k-k_*}{\Delta})\ln(k \sinh(2\pi k))\\ 
    &= \frac{2}{\Delta} \int \dd k\,f\qty(\frac{k-k_*}{\Delta})\qty(2\pi k + \ln(k)- \ln 2 + \ln(1-e^{-4\pi k})),
\end{Align}
where we used a standard identity for the logarithm of the hyperbolic sine. By the definition of the profile $f$, the first piece simply gives $4\pi k_*$. We disregard the constant term and leave the discussion of the exponentially suppressed term for later.
We are left with the logarithm term
\begin{Align}
 \expval{\ln k} &= \frac{1}{\Delta}\int \dd k\, f\qty(\frac{k-k_*}{\Delta})\ln k\\
 &=
\ln k_*-\sum_{n\ge2}\frac{1}{n}\qty(-\frac{\Delta}{k_*})^{n}m_n\\
 &=\ln k_*-\frac{m_2}{2}\frac{\Delta^2}{k_*^2}+\frac{m_3}{3}\frac{\Delta^3}{k_*^3}+O\qty(\frac{\Delta^4}{k_*^4}),
\end{Align}
where $m_n=\int \dd u\,u^n f(u)$ are the moments of the profile, and the $n=1$ term vanishes because $f$ is centered.
Strictly speaking, the second line requires $f(u)=0$ for $u\le -k_*/\Delta$, so that $p$ is supported on $k>0$, 
and the series converges only if $f$ is supported in $[-a,a]$ with $a\,\Delta<k_*$. In that case all equalities above are exact. 
For profiles with tails, such as the Gaussian restricted to $k\ge0$ and renormalized, the series does not converge, but 
the last line still holds as an asymptotic expansion in $\Delta/k_*$, up to corrections of order $e^{-k_*^2/2\Delta^2}$ which are negligible 
in the regime of interest.
Putting everything together, the entanglement entropy of the edge modes is given by (up to a constant)
\begin{equation}\label{eq:Sedge}
    S_{\mathrm{edge}} = \frac{A_h}{4} + 2\ln\frac{A_h}{4} - \mu_2\qty(\frac{\delta A}{A_h})^2 + \frac{2\mu_3}{3}\qty(\frac{\delta A}{A_h})^3 + \mathcal{O}\qty(\frac{\delta A^4}{A_h^4}),
\end{equation}
where $A_h = 4\pi\rho_h^2/\ell_P^2$ is the area of the horizon in Planck units, 
and $\delta A = 16\pi \Delta$ is the width of the area distribution, so that the corrections are controlled by the 
relative quantum uncertainty $\delta A/A_h$.
All regulator dependence through 
$\ln(\mathrm{Vol}^2(G)/V_C)$ sits in the discarded constant. As a consequence every 
$k_*$-dependent term in \eqref{eq:Sedge} is independent of the cutoff. The first two terms are universal, in the sense that they do not depend on the particular state or distribution.
Finally, the last exponential term in \eqref{eq:Sedgefull} contributes
\begin{equation}
    2\expval{\ln(1-e^{-4\pi k})} \approx -2 e^{-\frac{A_h}{4}} \expval{e^{-4\pi \Delta u}},
\end{equation}
where the last factor is $e^{8\pi^2 \Delta^2}$ for a Gaussian profile. This term is therefore of order
$e^{-\frac{A_h}{4}}$ and constitutes a non-perturbative effect of quantum gravity.
The Bekenstein--Hawking area law and its quantum corrections emerging from the edge mode entanglement entropy
between the spacetime regions I and II constitutes one of the main result of this paper. We note that, if one does not require the continuity condition
\eqref{eq:rNr} (or the invariance under the left action \eqref{eq:leftinvariance} at the quantum level), we would only get half of the area in the first term of the entropy.
In that case, the one-sided Gauss constraints remain first class as in Yang--Mills theory, the canonical pair is carried by the fields rather than by the frames, and the frames are only entangled through their frame factors.
\par
To conclude this section, we comment on the assumption made in substituting 
the Plancherel measure of $\mathrm{SL}^+(2,\mathbb{R})$ into 
\eqref{eq:finalentropy}. The derivation of \eqref{eq:trrhoIn} relied on $G$ being a group acting unitarily. The reduced 
density matrix commutes with the action of $G$ and is therefore 
block diagonal in the Plancherel decomposition. The trace then involves the kernel at the identity which can be rewritten using the Plancherel formula. 
None of these steps carries over directly to the semigroup $\mathrm{SL}^+(2,\mathbb{R})$.
The derivation can nevertheless be adapted to the semigroup. The steps that require a group, namely Schur's lemma and the evaluation of the kernel at the identity, can be replaced by the orthogonality relation of matrix elements inside the positive principal series of $\mathrm{SL}^+(2,\mathbb{R})$.
Taking the glued state to be the double identity intertwiner, this reproduces the second expression in \eqref{eq:trrhoIn}, with $\mathrm{Vol}(G)$ replaced by the regularized volume of $\mathrm{SL}^+(2,\mathbb{R})$ and the Plancherel measure replaced by \eqref{eq:SLplusPlancherel}, as shown in Appendix~\ref{app:SLplus}. 
The remaining assumption is the 
definition of the gluing itself, which can no longer be obtained as a quotient, since the elements of the semigroup are not invertible.
\section{Conclusion}\label{sec:conclusion}
The nature of the Bekenstein--Hawking entropy of black holes has been a central
research question since its introduction~\cite{Bekenstein:1972tm}.
Since it contains both Newton's constant and Planck's constant, it is one of the
earliest results hinting at the interplay between general relativity and quantum
theory. Bekenstein first proposed an entropy proportional to the horizon
area~\cite{Bekenstein:1973ur}, and Hawking's semiclassical computation of black
hole radiation fixed its coefficient~\cite{Hawking:1975vcx}. The result was soon
recovered without reference to matter fields, from the Euclidean gravitational
path integral~\cite{Gibbons:1976ue}. It was later reinterpreted as entanglement
entropy across the horizon~\cite{Bombelli:1986rw,Susskind:1994sm}, as a Cardy
counting of states~\cite{Strominger:1997eq}, and holographically as a
special case of the Ryu--Takayanagi formula~\cite{Ryu:2006bv}.
The present work continues this line of ideas, generalizing the edge mode
derivations available for gauge formulations of lower-dimensional
gravity~\cite{Lin:2017uzr,Lin:2018xkj,Mertens:2022ujr} and connecting them to
recent algebraic approaches~\cite{Witten:2021unn,Chandrasekaran:2022eqq,Jensen:2023yxy}. We argued 
that the Bekenstein--Hawking entropy of a static four-dimensional black
hole can be understood as the entanglement entropy of edge modes on the
bifurcation surface, which arise from the requirement of gauge invariance when the region beyond
the horizon is traced out.\par  
The quantum reference frame is introduced, as in~\cite{Donnelly:2016auv}, to restore diffeomorphism invariance of the subregion phase space. However, its structure is not chosen. Its phase space, symmetry group and charges are dictated by the theory. The extension adds no physical degree of freedom to the glued system~\cite{Speranza:2022lxr}. It provides instead a framework in which different descriptions of the same reduced phase space appear as gauge choices. In the spherically symmetric
sector, the corner is a point, the edge modes phase space is the cotangent
bundle of $\mathrm{SL}(2,\mathbb{R})$, and its quantization is
$L^2(\mathrm{SL}(2,\mathbb{R}))$. Since the harmonic analysis of the special linear group
is known, we were able to construct the gravitational entangling product of Donnelly and Freidel explicitly.
The key step of the construction is the gluing. In Yang--Mills theory, the
entangling product of~\cite{Donnelly:2016auv} matches the charges of the two
regions, which amounts to a quotient by the diagonal surface symmetries. 
Gauge transformations remain gauge independently on each side, since the matching only involves dressed 
charges, which are invariant under them. In gravity, the glued spacetime has a single smooth geometry at 
the corner. The corner charges of the fields of the two regions, which are tensors on the same tangent plane, 
must then agree, and this makes the one-sided Gauss constraints second class, except for the boost that 
preserves the corner geometry. Only the diagonal combination of the left actions, and this boost, 
remain gauge.
The
gravitational gluing therefore imposes two conditions, the matching of physical charges
and the invariance under the diagonal gauge action or continuity. The relative frame of
the two regions then survives as a physical degree of freedom. This has two
consequences. First, in the chosen gauge, the relative frame reduces to a single canonical pair: the
area and the relative rapidity. Second, each gluing condition contributes one
power of the Plancherel density to the entanglement entropy.
The entangling product of~\cite{Donnelly:2016auv}, which only matches the
charges, maximally entangles one of the two factors in the Plancherel
decomposition of $L^2(\mathrm{SL}(2,\mathbb{R}))$ across the corner, and
guarantees one power of the Plancherel density. The invariance
under the diagonal gauge action maximally entangles the other factor as well,
and it is this second power that brings the edge term to the full
Bekenstein--Hawking entropy.
\par 
It is remarkable that this pair is precisely the canonical structure of
Carlip and Teitelboim~\cite{Carlip:1993sa}, the relative rapidity playing the
role of their Euclidean opening angle at the bifurcation surface. It survives
the gluing for two reasons. Matching the charges cannot fix it, since the boost
of the normal plane leaves the corner geometry invariant. Nor is it removed by the one-sided boost, which remains gauge, since this boost also shifts the relative Killing time of the two exteriors and only the combination is physical. Fixing the latter leaves the relative rapidity.
That exactly one pair survives could have been predicted from Birkhoff's theorem, which makes the reduced phase space of spherically symmetric vacuum gravity two-dimensional. 
This turns the counting into a consistency check. Matching the charges alone leaves a six-dimensional phase space, while the gravitational gluing proposed here leaves exactly one pair, in which the frames carry the variable conjugate to the area. Gluing with independent left actions would also leave one pair, but carried by the fields, and the frames would then only account for half of the area.
The surviving pair also has a natural interpretation as a clock. 
For a stationary black hole, the one-sided boost at the corner extends to an isometry 
of each exterior, and the relative rapidity $\eta$ corresponds to the relative Killing 
time $T=\eta/\kappa$ between the two exteriors, where $\kappa$ is the surface 
gravity. Through the first law, the pair (area, relative rapidity) is then canonically equivalent to the pair (mass, difference of asymptotic Killing times) that Kucha\v{r} found as the reduced phase space of spherically symmetric vacuum gravity~\cite{Kuchar:1994zk}. 
What is new here is not this equivalence but the location of the pair. The gluing places it in the relative orientation of the two edge frames at the corner, where $\mathrm{SL}(2,\mathbb{R})$ acts on it, rather than in boundary terms added at the horizon or at infinity.
Kucha\v{r}'s time is the 
relative reading of two clocks at the two asymptotic infinities, introduced by 
parametrizing the asymptotic boundary terms. Here the same variable is the relative 
orientation of the two edge frames at the bifurcation point, introduced by 
diffeomorphism invariance at the corner. In both cases, time is a relation between two 
boundary frames, one for each exterior, demanded by the consistency of the variational 
principle, in the spirit of~\cite{Regge:1974zd,Donnelly:2016auv}.
Moreover, the restriction of the Hartle--Hawking state to
one exterior is thermal with respect to the boost~\cite{Bisognano:1976za,Kay:1988mu}, so
that the relative rapidity is a clock for the modular flow. This is the clock
added in the crossed product construction~\cite{Witten:2021unn,Chandrasekaran:2022eqq},
and the time it measures is the thermal time of Connes and
Rovelli~\cite{Connes:1994hv}. In this language, the area term of the entropy
is the product of the period of this clock in imaginary time and its conjugate
charge, which is the Lorentzian counterpart of the
statement of Carlip and Teitelboim that the entropy follows from fixing the
opening angle to $2\pi$.
Pairwise, these identifications are not new~\cite{Carlip:1993sa,Louko:1994tv,Harlow:2018tqv,Bousso:2020yxi,Chandrasekaran:2022eqq}. What the present construction adds is that all of them follow from the gluing of the two edge frames, together with the symmetry group whose Plancherel measure yields the area term.
The present construction thus connects within a single framework the edge
modes of Donnelly and Freidel, the canonical pair of
Carlip and Teitelboim, the time of
Kucha\v{r}, and the clock of the crossed product
construction~\cite{Witten:2021unn,Chandrasekaran:2022eqq}, all of which appear
as different descriptions of the same relative boost between the two edge
frames.
\par    
The area term arises as the leading order term in the edge mode contribution to the entanglement
entropy between the two exterior regions. Since we worked with vacuum gravity, the entropy computed here is that of
all degrees of freedom of region~I. In the
language of~\cite{DeVuyst:2024khu,DeVuyst:2024fxc}, it is the entropy relative
to the full set of quantum reference frames associated with the subregion,
which is the observer-independent quantity proposed there as a natural
definition of horizon entropy. The gauge freedom discussed in Section~\ref{sec:quantumandreferenceframe} gives a concrete instance of this observer dependence. The gauges $\tau=0$ and $\eta=0$ describe the same reduced phase space, but refer the variable conjugate to the area to different frames, the edge frames at the corner in the first and the asymptotic clocks in the second. They therefore lead to different factorizations across $S$, and only in the first does the entanglement of the frames account for the full area term.
The present construction also realizes their
proposal to use dynamical edge modes as quantum reference frames, with the
corner group $\mathrm{SL}(2,\mathbb{R})$ replacing the translation group of a
clock, and its Plancherel measure supplying the area term that the first law
supplies in the crossed product construction.\par 
It is also instructive to compare our result with the description of
Jackiw--Teitelboim gravity developed in
\cite{Blommaert:2018oro,Blommaert:2018iqz}. There, the authors propose that the gauge-theoretic description of the theory be based on the subsemigroup $\mathrm{SL}^+(2,\mathbb{R})$ rather than on $\spl{2}$, since its Plancherel measure 
reproduces the density of states of the Schwarzian theory, and they identify
the degrees of freedom frozen at the horizon with the black hole states. The
same semigroup and the same measure appear in our construction, but they
enter in a different way. First, in \cite{Blommaert:2018iqz} the restriction
to the semigroup is motivated by the density of states, by the absence of
singular geometries and by the classical limit of the quantum group
structure of three-dimensional gravity. Here it follows from a single
requirement on the edge mode, namely that the frame maps the future light
cone of the reference plane into that of the spacetime. Second, the origin of the group
itself is different. In that proposal it plays the role of the gauge group of Jackiw--Teitelboim gravity, while in our case it is the group of linear
transformations of the normal plane of the corner, which requires no gauge
theory formulation and is therefore available for the theory describing
spherically symmetric black holes.
Additionally, the identification of the representation label differs.
It is related to the Bekenstein--Hawking entropy by $2\pi k=S_{\mathrm{BH}}$ in the first
case and by $4\pi k=S_{\mathrm{BH}}$ in ours. Accordingly, a single power of
the Plancherel density gives the area term in Jackiw--Teitelboim
gravity~\cite{Lin:2018xkj}, while here one power comes from each of the two
gluing conditions. The two descriptions thus agree on the area term but not
on the logarithmic correction of the edge term, which is $\ln k$ in the
first case and $2\ln k$ in the second.
\par    
Several questions remain open. The most immediate one is to go beyond spherical
symmetry. For a general corner $S$, the edge modes transform under the full
corner symmetry group $\mathrm{Diff}(S)\ltimes\mathrm{SL}(2,\mathbb{R})^S$ of
Donnelly and Freidel~\cite{Donnelly:2016auv}, whose complete representation theory is
not known. A natural strategy is to describe the local $\mathrm{SL}(2,\mathbb{R})$
frames on $S$ by a Wess--Zumino--Witten type model on the corner, in which the
point-wise construction of the present work would provide the local area
density and the action of $\mathrm{Diff}(S)$ would be implemented as a
constraint on the glued states. Whether the area law, and in particular its
coefficient, survives the inclusion of $\mathrm{Diff}(S)$ is a nontrivial test
of the construction. A second question concerns the symmetries that move the
corner, which were excluded here by restricting to transformations that
preserve its position. Including them, as in the extended corner symmetry
group~\cite{Ciambelli:2021vnn}, is necessary to describe dynamical entangling
surfaces, and would clarify the relation with approaches based on the
deformations of the horizon~\cite{Carlip:1994gy}. It would also allow one to
address non-stationary situations, such as black holes formed by collapse,
where no bifurcation surface exists and the role of the second frame is
unclear. Finally, it would be interesting to couple matter fields, which could be done in two steps. 
Charged black holes provide a first test. For spherically symmetric Einstein--Maxwell 
theory, the reduced phase space contains an additional canonical pair for the electric charge
charge~\cite{Louko:1996dw}, so the gluing should leave the electric charge and a conjugate variable in addition to the area and the 
relative rapidity. In the present 
framework, the latter should be the relative $\mathrm{U}(1)$ frame of the Maxwell 
edge modes of the two regions, the analogue of the relative rapidity for the internal 
gauge group. Since the Plancherel measure of an abelian group is flat, these edge 
modes would contribute no area-like term, but they would test whether the 
gravitational and gauge edge modes glue consistently at the same corner. 
Local matter, on the other hand, evades Birkhoff's theorem and gives the bulk 
its own degrees of freedom. The entanglement entropy should then reproduce the full generalized entropy, which would allow a direct comparison with the crossed product construction beyond the vacuum case.
\section*{Acknowledgements}
I would like to thank Giulio Neri and Jerzy Kowalski-Glikman for insightful discussions and helpful comments on the manuscript.
I also benefited from discussions with
Matteo Bruno, Simone Speziale and Wolfgang Wieland. This work was supported by Grant~$\sharp63683$ for the
\href{https://withoutspacetime.org}{WithOut SpaceTime (WOST) project}
and Grant~$\sharp63132$ as part of the \href{https://eff.cstq.org}{Enrico Fermi
Fellowship program}, both funded by the John Templeton Foundation.

\appendix 
\section{Constraint Analysis of the Gluing}\label{app:constraints}
In this appendix, we derive the gluing conditions \eqref{eq:rightinvariance} and \eqref{eq:leftinvariance} from the Gauss laws \eqref{eq:gausslaw} and the continuity condition \eqref{eq:continuity}, and show that the one-sided boost remains gauge.
\par
For each region $\mathcal R\in\{\rm I,\rm II\}$, the kinematical phase space consists of the fields $\varphi_{\mathcal R}$, with presymplectic form $\Omega_{\mathcal R}$, and of the frame $e_{\mathcal R}$ with an independent momentum $p_{\mathcal R}$, with potential $\sigma^{\mathcal R}_S\langle p_{\mathcal R},e_{\mathcal R}^{-1}\delta e_{\mathcal R}\rangle$. The glued system is subject to the constraints
\begin{equation}\label{eq:constraints}
    \mathcal G_{\mathcal R}=N[\varphi_{\mathcal R}]\eval_S-e_{\mathcal R}\,p_{\mathcal R}\,e_{\mathcal R}^{-1}\approx0,
    \qquad
    \mathcal P\defeq p_{\rm I}-p_{\rm II}\approx0,
    \qquad
    \mathcal C=N[\varphi_{\rm I}]\eval_S-N[\varphi_{\rm II}]\eval_S\approx0 .
\end{equation}
The Gauss laws $\mathcal G_{\mathcal R}$ are \eqref{eq:gausslaw}, $\mathcal P$ is the matching of the charges of the frame reorientations, and $\mathcal C$ is the continuity condition \eqref{eq:continuity}. For $\lambda\in\mathfrak{sl}(2,\mathbb R)$, $\sigma^{\mathcal R}_S\langle\mathcal G_{\mathcal R},\lambda\rangle$ generates the corner-preserving diffeomorphism of region $\mathcal R$ with Jacobian $\mathds{1}+\lambda$ at $S$, acting on the fields and on the frame, and $\sigma^{\mathcal R}_S\langle p_{\mathcal R},\lambda\rangle$ generates the frame reorientations. Since $\sigma^{\rm I}_S=-\sigma^{\rm II}_S$, $\sigma^{\rm I}_S\langle\mathcal P,\lambda\rangle$ and $\sigma^{\rm I}_S\langle\mathcal C,\lambda\rangle=\sum_{\mathcal R}\sigma^{\mathcal R}_S\langle N[\varphi_{\mathcal R}],\lambda\rangle$ are the generators of the diagonal frame reorientations and of the diagonal diffeomorphisms acting on the fields only, respectively.
\par
Each family of constraints in \eqref{eq:constraints} is the moment map of a group action and closes on itself. The Gauss laws of the two regions commute, the left and right translations of a frame commute, and the frames do not enter $\mathcal C$. The only brackets that do not vanish weakly are therefore those between the Gauss laws and the continuity condition. A diffeomorphism of region $\mathcal R$ conjugates $N[\varphi_{\mathcal R}]$ and leaves the fields of the other region untouched, so that
\begin{equation}\label{eq:GCbracket}
    \{\sigma^{\rm I}_S\langle\mathcal G_{\rm I},\lambda\rangle,\langle\mathcal C,\mu\rangle\}
    =-\{\sigma^{\rm II}_S\langle\mathcal G_{\rm II},\lambda\rangle,\langle\mathcal C,\mu\rangle\}
    \approx\pm\langle N,[\lambda,\mu]\rangle ,
    \qquad \lambda,\mu\in\mathfrak{sl}(2,\mathbb R),
\end{equation}
where $N$ is the common value of the field charges and the overall sign depends on conventions. The diagonal Gauss law $\sum_{\mathcal R}\sigma^{\mathcal R}_S\langle\mathcal G_{\mathcal R},\lambda\rangle$ therefore commutes weakly with $\mathcal C$, since it rotates the field charges of both regions in the same way.
\par
To see which one-sided transformations survive, we use the basis of $\mathfrak{sl}(2,\mathbb R)$ adapted to $N=\Phi_hJ$, consisting of the boost generator $J$ and of
\begin{equation}
    K=\mqty(1&0\\0&-1),\qquad E=\mqty(0&1\\-1&0),
\end{equation}
which generate squeezes and rotations of the normal plane. The boost commutes with $N$, while $[J,K]=-2E$ and $[J,E]=-2K$. Since $\langle N,[\lambda,\mu]\rangle=\langle[N,\lambda],\mu\rangle$, the only non-vanishing brackets \eqref{eq:GCbracket} are
\begin{equation}
    \{\sigma^{\rm I}_S\langle\mathcal G_{\rm I},K\rangle,\langle\mathcal C,E\rangle\}
    =-\{\sigma^{\rm I}_S\langle\mathcal G_{\rm I},E\rangle,\langle\mathcal C,K\rangle\}
    \approx\pm4\Phi_h .
\end{equation}
In words, a squeeze or a rotation of region $\rm I$ alone changes $N[\varphi_{\rm I}]$ but not $N[\varphi_{\rm II}]$, and therefore violates continuity, while a boost leaves $N$ invariant. The constraints thus split as follows.
\begin{itemize}
    \item Second class: the one-sided squeeze and rotation, $\langle\mathcal G_{\rm I},K\rangle$ and $\langle\mathcal G_{\rm I},E\rangle$, together with the corresponding components of the continuity condition, $\langle\mathcal C,E\rangle$ and $\langle\mathcal C,K\rangle$. Each continuity condition acts as a gauge fixing of one of these one-sided transformations, and these four constraints are solved rather than quotiented.
    \item First class: all the other constraints, which generate gauge transformations. These are the diagonal Gauss law, i.e.\ the diffeomorphisms of the glued spacetime, which act on both regions with the same Jacobian at $S$, the one-sided boost $\langle\mathcal G_{\rm I},J\rangle$, the matching $\mathcal P$, i.e.\ the diagonal frame reorientations, and $\langle\mathcal C,J\rangle=2(\Phi^{\rm I}-\Phi^{\rm II})\eval_S$, which states that the two regions have the same area.
\end{itemize}
\par
Solving the second-class constraints identifies the field charges with the left momenta of the frames, $N[\varphi_{\rm I}]\eval_S=N[\varphi_{\rm II}]\eval_S=e_{\rm I}p_{\rm I}e_{\rm I}^{-1}=e_{\rm II}p_{\rm II}e_{\rm II}^{-1}$. The frame variables commute with $\mathcal C$, and the matrix of brackets of the second-class constraints only pairs the one-sided Gauss law with $\mathcal C$. Every correction in the Dirac bracket of two frame variables therefore contains a bracket of a frame variable with $\mathcal C$, and the Dirac brackets of the frames are the brackets of $T^*G\times T^*G$. On $\mathcal C=0$, the field part of the diagonal Gauss law, $\sigma^{\rm I}_S\langle\mathcal C,\lambda\rangle$, vanishes, and the diagonal Gauss law only generates the diagonal left translation of the frames. Together with $\mathcal P$, its quantization gives \eqref{eq:rightinvariance} and \eqref{eq:leftinvariance}.
\par
The one-sided Gauss law along $N$ generates a diffeomorphism of region $\rm I$ whose Jacobian at $S$ is a boost of the normal plane. It can be chosen to coincide with the Killing boost of rapidity $\beta$ near $S$ and to vanish far from $S$. Near $S$, it leaves the fields invariant and rotates the frame, $\eta\mapsto\eta+\beta$. Where it is cut off, it deforms the fields of region $\rm I$. Let $\tau$ be the sum over the two regions of the future-directed Killing time, normalized at the asymptotic end, elapsed along $\Sigma_{\mathcal R}$ from the surface of constant Killing time tangent to $\Sigma_{\mathcal R}$ at $S$ to the asymptotic end. It vanishes for the straight slice of Figure~\ref{fig:blackhole}, is invariant under diffeomorphisms that are smooth at $S$ and vanish at infinity, and is the variable conjugate to the mass of~\cite{Kuchar:1994zk}. The deformation shifts it by $\beta/\kappa$, so that only $\eta-\kappa\tau$ is gauge invariant, up to orientation conventions. The field part alone is not gauge, since by \eqref{eq:Iomega} its charge with respect to $\Omega_{\rm I}+\Omega_{\rm II}$ is $\sigma^{\rm I}_S\langle N,\lambda\rangle\neq0$.
We use the one-sided boost to set $\tau=0$, and the diffeomorphisms that are smooth at $S$ to bring the fields to the Kruskal form of Section~\ref{sec:dimensionallyreducedblackhole}, with $N=\Phi_hJ$. The fields are then the Kruskal solutions on the straight slice of Figure~\ref{fig:blackhole}, a family labelled by a single parameter, on which the two-form $\Omega_{\rm I}+\Omega_{\rm II}$ vanishes. The remaining variables are the frames, subject to \eqref{eq:rightinvariance} and \eqref{eq:leftinvariance}, with the symplectic form \eqref{eq:gluedOmega}.
\par
In the gluing of~\cite{Donnelly:2016auv}, the matching condition only involves the dressed charges, which are invariant under the gauge transformations of either region. There is no analogue of $\mathcal C$, all one-sided Gauss constraints remain first class, and the relative trivialisation of the two regions is pure gauge. In the spherically symmetric theory, the same would hold without \eqref{eq:continuity}. The pair would then be carried by $\tau$ alone, and the frames would only be entangled through their frame factors.
\section{Entanglement Entropy on the Semigroup \texorpdfstring{$\mathrm{SL}^+(2,\mathbb{R})$}{SL+(2,R)}}\label{app:SLplus}

In section~\ref{sec:entanglemententropy}, the trace \eqref{eq:trrhoIn} was obtained using three properties of the group $G$: the reduced density matrix is block diagonal by Schur's lemma, its kernel only depends on $gg'^{-1}$, and the trace reduces to the kernel at the identity, evaluated with the Plancherel formula. None of these is available for $\mathrm{SL}^+(2,\mathbb{R})$, which contains neither inverses nor the identity. In this appendix, we show that the same result follows from a single property of the semigroup, the orthogonality of the matrix elements of its positive principal series.

\par 
Let $U_k$ be the positive principal series representation of $\mathrm{SL}^+(2,\mathbb{R})$ on $\mathcal{P}_k$, realized on functions of $x>0$ as in section~\ref{sec:entanglemententropy}, and let $\{\ket{k;a}\}$ be a basis of $\mathcal{P}_k$. Its matrix elements are the functions on the semigroup
\begin{equation}
    K^k_{ab}(g)\defeq\bra{k;a}U_k(g)\ket{k;b},\qquad g\in\mathrm{SL}^+(2,\mathbb{R}).
\end{equation}
As in the group case, the frame index $b$ is acted on by right multiplication, and the spacetime index $a$ by left multiplication. If the basis is continuous, sums over $a,b$ are understood as integrals and $\delta_{ab}$ as a Dirac delta. We write $\dim\mathcal{P}_k=\sum_a\delta_{aa}$, which is divergent and regularized as in section~\ref{sec:entanglemententropy}.

Following~\cite{Blommaert:2018oro}, we use the hyperbolic basis $\{\ket{k;s}\}_{s\in\mathbb{R}}$ of eigenvectors of the boost generator, normalized as $\braket{k;s}{k;s'}=\delta(s-s')$, so that the indices $a,b$ are continuous labels $s\in\mathbb{R}$ and sums over them are integrals.\footnote{The parabolic eigenvectors, which form a basis for $\mathrm{SL}(2,\mathbb{R})$, do not form a basis on $x>0$~\cite{Blommaert:2018oro}.} We denote by $\dd g$ the restriction to $\mathrm{SL}^+(2,\mathbb{R})$ of the Haar measure of $\mathrm{SL}(2,\mathbb{R})$, i.e.\ the volume form of the metric $\tfrac12\Tr(g^{-1}\dd g)^2$. As is discussed in~\cite{Blommaert:2018oro}, the $q\to1$ limit of the Plancherel decomposition of the quantum semigroup $\mathrm{SL}^+_q(2,\mathbb{R})$ gives
\begin{equation}
    L^2(\mathrm{SL}^+(2,\mathbb{R}))\simeq\int_0^\infty\dd\mu(k)\;\mathcal{P}_k\otimes\mathcal{P}_k^*,\qquad \dd\mu(k)=k\sinh(2\pi k)\,\dd k,
\end{equation}
and the functions $\sqrt{\mu(k)}\,K^k_{ab}(g)$ form a complete orthonormal set. Orthonormality reads
\begin{equation}\label{eq:SLplusorthogonality}
    \int_{\mathrm{SL}^+}\dd g\,K^k_{ab}(g)\,\overline{K^{k'}_{cd}(g)}=\frac{\delta(k-k')}{\mu(k)}\,\delta_{ac}\delta_{bd}.
\end{equation}
This orthogonality relation is the only property of $\mathrm{SL}^+(2,\mathbb{R})$ used below. For a group, it is the standard orthogonality of matrix elements, which is equivalent to the Plancherel formula used in Section~\ref{sec:entanglemententropy}. An overall constant in the normalization of $\mu$ rescales the right-hand side, and, as already mentioned, only shifts the entropy by a constant.

\par  
For a group, the wavefunction of the doubly glued state \eqref{eq:doublygluedstate} contains characters of the relative frame. Unitarity allows one to write them as $\chi_k(g_Ig_{II}^{-1})=\sum_{a,b}U_{ab}(g_I)\overline{U_{ab}(g_{II})}$. On the semigroup, $g_{II}^{-1}$ is not causal, so that only the second form makes sense. We therefore define the glued state as the double identity intertwiner
\begin{equation}\label{eq:SLplusgluedstate}
    \Psi(g_I,g_{II})=\int\dd\mu(k)\,\psi(k)\sum_{a,b}K^k_{ab}(g_I)\,\overline{K^k_{ab}(g_{II})},\qquad g_I,g_{II}\in\mathrm{SL}^+(2,\mathbb{R}),
\end{equation}
which contracts the frame index $b$ and the spacetime index $a$ of the two regions, as in the group case. This is the assumption mentioned at the end of section~\ref{sec:entanglemententropy}: for a group, \eqref{eq:SLplusgluedstate} follows from the invariance under the diagonal actions, while here it is taken as the definition of the gluing.
\par    
Tracing over region II, the kernel of the reduced density matrix is
\begin{Align}
    \rho_{\rm I}(g,g')&=\int_{\mathrm{SL}^+}\dd g_{\mathrm{II}}\,\Psi(g,g_{\mathrm{II}})\,\overline{\Psi(g',g_{\mathrm{II}})}\\
    &=\int\dd\mu(k)\,\dd\mu(k')\,\psi(k)\overline{\psi(k')}\sum_{a,b,c,d}K^k_{ab}(g)\,\overline{K^{k'}_{cd}(g')}\int_{\mathrm{SL}^+}\dd u\,\overline{K^k_{ab}(u)}\,K^{k'}_{cd}(u).
\end{Align}
The last integral is the complex conjugate of \eqref{eq:SLplusorthogonality}, i.e.\ $\delta(k-k')\delta_{ac}\delta_{bd}/\mu(k)$. We can thus write
\begin{equation}\label{eq:SLplusrho}
\rho_{\rm I}(g,g')=\int\dd\mu(k)\,\abs{\psi(k)}^2\,E_k(g,g'),\qquad E_k(g,g')\defeq\sum_{a,b}K^k_{ab}(g)\,\overline{K^k_{ab}(g')}.
\end{equation}
Using \eqref{eq:SLplusorthogonality} once more, the kernels $E_k$ compose as
\begin{equation}\label{eq:SLplusprojector}
    \int_{\mathrm{SL}^+}\dd u\,E_k(g,u)\,E_{k'}(u,g')=\frac{\delta(k-k')}{\mu(k)}\,E_k(g,g').
\end{equation}
Up to normalization, $E_k$ is therefore the projector on the sector $k$ of $L^2(\mathrm{SL}^+(2,\mathbb{R}))$, and \eqref{eq:SLplusrho} states that $\rho_{\rm I}$ is proportional to the identity in each sector. In other words, both the frame factor and the spacetime factor are maximally mixed. This is the result that Schur's lemma provided in the group case, now obtained without using inverses.

\par    
Applying \eqref{eq:SLplusprojector} $n-1$ times, the kernel of $\rho_{\rm I}^n$ is
\begin{equation}
    \rho_{\rm I}^n(g,g')=\int\dd\mu(k)\,\abs{\psi(k)}^{2n}\,E_k(g,g').
\end{equation}
Its trace is obtained by setting $g'=g$ and integrating over the semigroup. Using \eqref{eq:SLplusorthogonality} with $k'=k$, $c=a$ and $d=b$,
\begin{equation}
    \int_{\mathrm{SL}^+}\dd g\,E_k(g,g)=\sum_{a,b}\int_{\mathrm{SL}^+}\dd g\,\abs{K^k_{ab}(g)}^2=\frac{\delta(0)}{\mu(k)}\sum_{a,b}\delta_{aa}\delta_{bb}=\frac{\delta(0)}{\mu(k)}\,\qty(\dim\mathcal{P}_k)^2,
\end{equation}
where $\delta(0)$ denotes $\delta(k-k')$ at $k'=k$. Since $\dd\mu(k)=\mu(k)\dd k$, the factors of $\mu$ cancel and
\begin{equation}\label{eq:SLplustrace}
    \Tr\rho_{\rm I}^n=\int\dd k\;\delta(0)\,\qty(\dim\mathcal{P}_k)^2\,\abs{\psi(k)}^{2n}.
\end{equation}
Note that we never evaluated a kernel at the identity, which does not belong to $\mathrm{SL}^+(2,\mathbb{R})$.
Equation \eqref{eq:SLplustrace} contains the two divergent factors $\delta(0)$ and $\dim\mathcal{P}_k$. For a group, the same computation can be done using the orthogonality of matrix elements, and must agree with \eqref{eq:trrhoIn}. Comparing \eqref{eq:SLplustrace} with the second expression in \eqref{eq:trrhoIn}, $\mathrm{Vol}(G)\int\dd k\,\mu(k)\dim\mathcal{P}_k\abs{\psi(k)}^{2n}$, fixes
\begin{equation}
    \delta(0)=\frac{\mathrm{Vol}(G)\,\mu(k)}{\dim\mathcal{P}_k}=\frac{V_C}{2\pi},
\end{equation}
where the second equality uses the regularized dimension \eqref{eq:Dregularization}. In other words, $\delta(0)$ is the inverse of the spacing $2\pi/V_C$ between the labels $k$ once the Cartan subgroup is regularized to volume $V_C$. Assuming the same regularization for the semigroup, with $\mathrm{Vol}(G)$ replaced by the regularized volume of $\mathrm{SL}^+(2,\mathbb{R})$, \eqref{eq:SLplustrace} becomes
\begin{equation}
\Tr\rho_{\rm I}^{n}=\frac{V_C}{2\pi}\qty(\frac{2\pi\,\mathrm{Vol}(\mathrm{SL}^+)}{V_C})^2\int\dd k\,\mu(k)^2\abs{\psi(k)}^{2n}=\frac{2\pi\,\mathrm{Vol}(\mathrm{SL}^+)^2}{V_C}\int\dd k\,\mu(k)^2\abs{\psi(k)}^{2n},
\end{equation}
which is equation \eqref{eq:410} with a single series and the Plancherel density \eqref{eq:SLplusPlancherel}. In particular, $n=1$ gives the normalization $\int\dd k\,p(k)=1$ of the state, and the entropy \eqref{eq:finalentropy} follows as in the main text.

\bibliographystyle{JHEP}
\bibliography{refs}

\end{document}